\documentclass[a4paper,fleqn]{cas-dc}

\usepackage[numbers]{natbib}
\usepackage{pifont}

\usepackage{graphicx} 
\usepackage{booktabs} 
\usepackage{array}    
\usepackage[table]{xcolor} 
\usepackage{caption}
\usepackage{enumitem}
\usepackage{amsmath,bm}
\usepackage{float}
\usepackage{comment}
\usepackage{enumitem}

\def\tsc#1{\csdef{#1}{\textsc{\lowercase{#1}}\xspace}}
\tsc{WGM}
\tsc{QE}
\tsc{QE}
\tsc{EP}
\tsc{PMS}
\tsc{BEC}
\tsc{DE}

\begin{document}

\begin{titlepage}
    \noindent {\LARGE \textbf{Title Page}} \\
    \rule{\textwidth}{0.4pt} \\ 
    \vspace{0.75cm}
    
    \noindent \textbf{Article title:} \\
    {\Large Motile Bacteria Modify Salt Precipitation Patterns in Dried Sessile Droplet} \\
    \vspace{1cm} 
    
    \noindent \textbf{Author names:} \\
    {\normalsize Yumeng Zhao \textsuperscript{a,b,*}, Boyoung Jeong \textsuperscript{a,c}, Markus C. Noll \textsuperscript{d}, and Sheng C. Dai \textsuperscript{a}} \\
    \vspace{0.75cm}
    
    \noindent \textbf{Affiliations and address:} \\
    {\small
    \textsuperscript{a} \textit{School of Civil and Environmental Engineering, Georgia Institute of Technology, 790 Atlantic Drive, Atlanta, GA 30332, USA} \\
    \textsuperscript{b} \textit{Department of Civil and Environmental Engineering, University of Nebraska - Lincoln, 1110 S 67th Street, Omaha, NE 68182, USA} \\
    \textsuperscript{c} \textit{Department of Civil, Architectural, and Environmental Engineering, Illinois Institute of Technology, 3201 S Dearborn Street, Chicago, IL 60616, USA}
        \\
        \textsuperscript{d} \textit{Blue Valley North High School, 12404 Lamar Avenue, Overland Park, KS 66209, USA}
    \vspace{1.5cm}
    }
    
    \noindent \textbf{* Corresponding Author:} \\
    Yumeng Zhao \\
    Email: \href{mailto:yzhao52@unl.edu}{yzhao52@unl.edu}
    
\end{titlepage}

\let\WriteBookmarks\relax
\shorttitle{Motile Bacteria Modify Salt Precipitation Patterns}    
\shortauthors{Zhao et al.}  

\title [mode = title]{Motile Bacteria Modify Salt Precipitation Patterns in Dried Sessile Droplet}  


\author[1,2]{Yumeng Zhao}[orcid=0000-0002-3828-178X]\cormark[1]
\credit{Conceptualization, Methodology, Formal analysis, Data curation, Funding acquisition, Supervision, Writing - original draft, Writing - review \& editing}
\affiliation[1]{organization={School of Civil and Environmental Engineering, Georgia Institute of Technology}, addressline={790 Atlantic Drive}, city={Atlanta}, postcode={30332}, state={GA}, country={USA}}
\affiliation[2]{organization={Department of Civil and Environmental Engineering, University of Nebraska--Lincoln},addressline={1110 S 67th Street},city={Omaha},postcode={68182},state={NE},country={USA}}

\author[1,3]{Boyoung Jeong}
\credit{Methodology, Writing - review \& editing}
\affiliation[3]{organization={Department of Civil, Architectural, and Environmental Engineering, Illinois Institute of Technology, 3201 S Dearborn Street}, city={Chicago},postcode={60616},state={IL},country={USA}}

\author[4]{Markus C. Noll}
\credit{Data curation}
\affiliation[4]{organization={Blue Valley North High School},addressline={12404 Lamar Avenue},city={Overland Park},postcode={66209},state={KS},country={USA}}

\author[1]{Sheng C. Dai}[orcid=0000-0003-0221-3993]
\credit{Conceptualization, Methodology, Data curation, Funding acquisition, Supervision, Writing - review \& editing}
\affiliation[1]{organization={School of Civil and Environmental Engineering, Georgia Institute of Technology},addressline={790 Atlantic Drive},city={Atlanta},postcode={30332},state={GA},country={USA}}

\cortext[1]{Corresponding author}
\ead{yzhao52@unl.edu}


\begin{abstract}
Motile \textit{Escherichia coli} bacteria can alter salt crystallization patterns during the evaporation of sessile droplets. In dilute bacterial suspensions in deionized water, dried bacteria cells predominantly accumulate at the droplet periphery, consistent with the classic "coffee-ring" effect. At higher cell densities, however, the bacterial distribution becomes more uniform. In the absence of bacteria, pure Phosphate Buffered Saline also forms salt crystals in a coffee-ring pattern. When bacteria are present alongside the salt solute, additional isolated crystals appear near the droplet center, with their abundance increasing with bacterial concentration, while crystals at the periphery adopt dendritic morphologies that extend radially. To investigate these phenomena, we used a Stokes-based analytical model to estimate the evolution of internal flow fields and compare them with bacterial motility. Then a finite volume model is implemented for bacteria and salt transport and adsorption, and a stochastic model for salt nucleation was developed, which successfully explains the crystallization pattern seen in the experiments. Our results show that bacterial motility can overcome evaporation induced flow during early stage, enabling bacteria cells to serve as nucleation sites and thereby altering the final crystalline morphology. This work highlights the potential of motile microorganisms to actively control evaporative crystallization, with implications for porous media flow and microfluidic deposition processes.
\end{abstract}





\maketitle
\section{Introduction}
When a sessile droplet contains solutes or suspended colloids particles and  is evaporated under a pinned triple-contact line condition, it typically leaves behind a dense, dendritic precipitate ring at its perimeter, and a relatively clean interior. This ubiquitous phenomenon, famously known as the "coffee-ring effect," is driven by a strong, capillary-induced, radial outward flow~\citep{deegan1997capillary,deegan2000contact}. The internal migration of the particles compensates for the solvent lost to evaporation at the droplet's edge, dictated by the geometrical constraint of maintaining an equilibrium shape with a fixed contact line~\citep{kaya2010pattern,kumar2022patterns}. This characteristic deposition can evolve into more complex patterns when the internal flow is no longer exclusively governed by the evaporation flux~\citep{shahidzadeh2008salt,parsa2018mechanisms}. For instance, the introduction of surfactants induces Marangoni eddies that reverse the internal fluid velocity field~\citep{hu2006marangoni,sempels2013auto}. Similarly, thermal gradients can transform the standard ring into "coffee-eyes"—a morphology featuring a thick central stain surrounded by a thin outer ring~\citep{li2015coffee}. Furthermore, shape-dependent capillary interactions between suspended particles and the droplet's free surface can completely suppress the coffee-ring formation~\cite{yunker2011suppression}.

The study and manipulation of evaporation-induced deposition patterns have gained significant interest across various fields. In the realm of biosensing, these patterns have been leveraged as a rapid, low-cost methodology for protein detection~\citep{wen2013coffee} and the diagnosis of blood cell diseases~\citep{chen2016blood,hertaeg2021pattern}. Beyond diagnostics, controlling this deposition is critical for electronic chip industry and nanotechnological applications, including achieving uniform coatings in inkjet printing and painting~\citep{yu2019inhibit,hu2020general}, guiding the complex assembly of nanoparticles~\citep{han2012learning}, developing novel particle separation strategies~\citep{wong2011nanochromatography} which is relevant to technological practices like coating technologies, fabrication of polymer
films, and microelectronics~\citep{kaya2010pattern}.

Beyond passive colloidal suspensions, the introduction of active particles—such as motile bacteria and chemically active colloids (e.g., Janus particles)—adds a highly dynamic layer of complexity to evaporative assembly~\citep{banik2026coffee}. Swimming bacteria, for instance, can produce non-coffee-ring deposition patterns due to the competing effects of their motility, volume fraction, evaporation rate, and capillary flow~\citep{wilting2025active}, with bacterial mobility playing a major role in determining the growth dynamics at the droplet edge~\citep{andac2019active}. Furthermore, adding motile bacteria can alter not only their own spatial distribution but also that of the background colloidal bath, thereby influencing the dried patterns of complex fluids such as whole blood~\citep{roy2025insights}.

While active and passive colloids typically form distinct particulate stains upon drying, salt solutions produce far more complex patterns. The final deposit is not only spatially inhomogeneous but also displays a rich variety of crystal morphologies, and the pattern is highly sensitive to nucleation processes. In a pinned evaporating droplet, patterns ranging from concentric circles to dendritic and lattice structures can emerge, depending on the evaporation rate and initial salt concentration~\citep{morinaga2018emergence}. The crystallization dynamics are governed by both the transport properties in the liquid (captured by the P\'{e}clet number) and the interfacial properties of the different crystalline phases~\citep{shahidzadeh2008salt}. The complexity increases when polymers are present: studies on drying drops of poly(styrene sulfonate) and sodium chloride have observed fractals, dendrites, periodic concentric rings, needle‑like crystals, and small triangular crystals~\citep{kaya2010pattern}. Moreover, environmental factors such as humidity, temperature, and pressure significantly influence the formation of these structures.

With respect to salt precipitation, studies have shown that motile bacteria can fundamentally alter crystal morphology, transforming simple passive precipitates into intricate, three-dimensional, or dendritic biosaline structures, with the final pattern depending critically on both the bacterial species and the specific salt type present~\citep{gomez2014drying,gomez2016rich,hegde2021vapor,mehdi2024evaporation}. In addition to their physical influence on fluid transport, bacteria actively participate in the chemical precipitation process through localized surface interactions. Most motile bacteria, such as \textit{Escherichia coli} (\textit{E.~coli}), possess a net negatively charged cell envelope, which electrostatically attracts positively charged metal cations and concentrates them at the cell–fluid interface~\citep{deng2010microbial,kenward2009precipitation,tourney2014role,petrash2017microbially}. Consequently, the bacterial surface serves as an active nucleation site. However, accurately predicting, and eventually, manipulating the final morphology and spatial arrangement of these salt crystals requires coupling these local nucleation kinetics with the macroscopic transport of both the motile bacteria and the dissolved ions within the dynamic evaporative flow field. Because the interplay between these multi-component dynamics has yet to be systematically investigated, it represents a significant gap in the study of complex fluid evaporation.

In this study, we experimentally investigate how motile \textit{E. coli} alters the precipitation patterns of Phosphate Buffered Saline (PBS) solution in evaporating sessile droplets with pinned contact lines. To isolate the underlying physical mechanisms, we compare these multi-component dynamics against the baseline evaporation of pure PBS and \textit{E. coli} in deionized (D.I.) water suspensions. Time-lapse imaging at the contact line and within the bulk fluid captures both the real-time bacterial motility and the final deposition morphology. Furthermore, we employ analytical and numerical approaches to compute the internal flow fields and the resulting transport of bacteria and dissolved salts. By incorporating the stochastic redistribution of concentrations into our final deposition model, we successfully replicate the experimentally observed precipitation patterns. Finally, we highlight the potential of using active particles to tune precipitation and crystallization strategies across diverse industrial applications.

\section{Materials and Methods}
\subsection{Experimental Methodology}
We chose \textit{Escherichia coli} (\textit{E. coli}, ATCC 9637) as the motile bacteria because their motility has been well documented and extensively studied in various applications~\citep{darnton2007torque,schwarz2016escherichia,jeong2020experimental,zhao2020impacts,jeong2026rheology}.
\textit{E. coli} were grown in a 1000 ml nutrient broth (NB) medium (ATCC 1376) at 37\textdegree C, 140 rpm and 1 atm for 18h. After the cultivation, the bacteria suspensions were centrifuged twice at 2000 rpm for 10 min in either PBS solution, which contains 8 g/L NaCl, 0.2 g/L KCl, 1.44 g/L Na$_2$HPO$_4$ and 0.24g/L KH$_2$PO$_4$ or in D.I. water. The cells were targeted to different concentration that was determined by both OD600 and standard colony-forming unit (cfu) counting. The zeta potential of the cells was determined to be -17.3$\pm$1.18 mV by Malvern Zetasizer Nano Z. After carefully cleaning the borosilicate chamber slides with ethanol and D.I. water and dried, \textit{E. coli} suspensions of about 2$\mu$L were transferred onto the slides with cover open to let the sessile droplets evaporate at ambient temperature of about 21\textdegree C. Using Zeiss LSM700/710 Inverted Confocal Microscope, we took photos at the contact lines at interval time and the stitched dried patterns. The total evaporation time was about 24-30 minutes. We also took some short videos at different position and depth of the droplets while they are drying to obtain bacteria motility in live. To quantify droplet evaporation rate, we conducted accompanying droplet evaporation tests under the same condition and used the ram\'{e}-hart tensiometer to measure the decaying of droplet contact angle and volume over time.

\subsection{Numerical Methodology}
We hypothesize that two factors contribute to the altered salt precipitation patterns by the presence of \textit{E.~coli}: i) the bacteria provide the nucleation sites and ii) the motility of bacteria further enhances the likelihood of salt precipitation away from the periphery. To test the hypotheses, we first extracted and analyzed bacteria swimming motion from videos, and obtained swimming velocity and apparent diffusivity. Then, we conducted a combined analytical, numerical simulation approach along with a stochastic salt precipitation modeling to establish the entire drying process with the presence of \textit{E.~coli} and PBS solute. We used MATLAB for the coding. The detailed theoretical derivations are presented in the Appendix~\ref{App_Hu} to~\ref{app_crystal} and the corresponding code is attached in the \nameref{Supp_Info} S5-S8. \nameref{Supp_Info} S1 provides a summary of all the rest supplementary documents. The modeling approach proceeds in three stages, described in turn below.

\subsubsection{Evaporation-Induced Flow Field}
We first computed the flow field inside the evaporating droplet following the analytical Stokes flow solution by Hu \& Larson, 2005~\citep{hu2005analysis}. Refer to the Appendix~\ref{App_Hu} for the equations used, Appendix~\ref{App_Flow} for the parameters specific to our experimental conditions, and the \nameref{Supp_Info} S5-S6 for the code implementation. The key assumptions of the model include a pinned contact line, axisymmetric flow, and quasi-steady water vapor diffusion in the surrounding ambient. The flow is considered to be in the low-Reynolds-number regime, such that inertial forces are neglected and the motion is governed by the Stokes equations. Gravity is also ignored, and consequently the droplet shape is determined solely by surface tension, resulting in a spherical-cap geometry; this simplification is justified under the experimental conditions, as the Bond number is much smaller than unity. To derive the analytical solution of the fluid field, the droplet profile is further approximated to have a parabolic shape, which becomes more accurate as the droplet evaporates out~\citep{hu2005analysis}. We assume the fluid flow is only driven by evaporation flux, and the presence of \textit{E.~coli} and PBS solute does not affect the flow field but rather follows the streamline.

\subsubsection{Transport and Deposition of \textit{E.~coli} and Salt}\label{sec:method_transport}
After the background fluid field is determined, we then used the explicit upwind Finite Volume Method (FVM) to calculate the transport of \textit{E.~coli} and PBS solute in the bulk. We first proved in the Appendix~\ref{App_ALE} that under the parabolic droplet profile approximation, the droplet shape can be transferred into a static profile under a special Arbitrary Lagrangian Eulerian (ALE) framework that significantly ease the computation. Second, initially uniformly distributed \textit{E.~coli} and PBS solute were assumed to flow along the background flow streamline with their own diffusivities (see the Appendix~\ref{app_Transport} and~\nameref{Supp_Info} S7).

The spatiotemporal distributions of \textit{E.~coli} and salt in the bulk, as well as their final deposition patterns, were modeled for an initial density of $c_E = 4 \times 10^8$ cfu/ml and $c_S = 0.1515~mol/L$. \textit{E.~coli} motility is included as an effective diffusivity of $D_E = 0.8\mu m^2/s$ estimated from experimental video data, see~\nameref{Supp_Info} S2 and the Results section~\ref{sec:motility} for details. To quantify the effect of motility on species transport, we compared this baseline to two alternative scenarios: an estimated diffusivity of $D_E = 0.3~\mu m^2/s$ for non-motile bacteria derived from the Stokes-Einstein equation~\citep{einstein1956investigations,berg2025random}, and an upper ceiling of $D_E = 8.0 ~\mu m^2/s$, representing the highly motile bacteria observed in our experiments. We applied zero-flux boundary conditions for both bacteria and salt at the free surface, along with a simple linear adsorption law for the bacteria at the substrate, while the PBS solute has no adsorption onto any surfaces considered in this stage. 
For the salt solute, we estimated a diffusivity of $D_S = 10^3 \mu m^2/s$. Given this high diffusivity and the correspondingly low P\'{e}clet number, we approximated a uniform salt concentration along the vertical axis of the droplet bulk, which significantly accelerated computation times. We further assumed that no salt nucleation occurs until the droplet has entirely dried out, which is consistent with our experimental observations where no salt precipitation was detected until the final seconds of evaporation. See Appendix~\ref{app_Transport} and \nameref{Supp_Info} S7 for detailed derivation and implementation.

\subsubsection{Stochastic Salt Crystallization Model}\label{sec:method_stoch}
Up to now, the model can only estimate the bacteria deposition along a 1$D$ axisymmetric substrate domain. Inspired by the idea of Stochastic Cellular Automata~\citep{weimar1994class}, we then introduced stochasticity detailed in the Appendix~\ref{app_crystal} and the \nameref{Supp_Info} S8 in order to determine the salt precipitation patterns. At the final stage of the transport computation, the depth-averaged bulk salt concentration is collapsed onto its radial positions on the substrate, and the deposited \textit{E~.coli} radial distribution on the substrate is also used. These 1$D$ radial profiles are mapped into 2$D$ axisymmetric areal-density fields on the circular substrate, designated $E_{base}$ and $S_{base}$. Because bulk fluid transport of salt is assumed to be uninfluenced by the presence of bacteria, $S_{base}$ remains the same across all cases, while $E_{base}$ is omitted for the control (pure salt) case.

While these base fields are assumed to follow deterministic fluid streamlines alongside intrinsic molecular diffusion, hydrodynamic disturbances and random bacterial motion introduce spatial heterogeneity in real systems. To capture these stochastic fluctuations, a randomized field layer is superimposed on the idealized $E_{base}$ and $S_{base}$ profiles, updating them to $E_{stoch}$ and $S_{stoch}$, respectively. For \textit{E.~coli}, stochasticity is modeled via a Gaussian-kernel-smoothed random field. For salt, an additional layer of stochasticity is introduced to incorporate the co-localization effect: bacteria provide favorable adsorption sites for solutes throughout the evaporation process but this is not considered in the aforementioned species transport process, so to compensate, the PBS solute concentration redistribution is collapsed into a single final step just before the droplet is completely dried. In this step, the stochastic salt concentration in the bulk is rearranged to account for the non-homogeneity introduced by~\textit{E.~coli}, and the updated final bulk salt concentration field is then projected onto the substrate.

Finally, $E_{stoch}$ and $S_{stoch}$ serve as input fields to compute the final salt crystallization morphology using a Stochastic Cellular Automata framework. We define a baseline absolute nucleation threshold $S_{crit}$, required for salt precipitation. To reflect the physical nucleation process, this threshold is locally modulated by two conditions: (i) high local \textit{E.~coli} concentrations reduce the energy barrier, facilitating heterogeneous nucleation; (ii) active nucleation at a given site promotes crystal growth into adjacent neighboring sites. Detailed descriptions of the algorithm, code implementation, and parameter selections are provided in Appendix~\ref{app_crystal} and Supplementary Information Section~S\ref{Supp_S8}. All the stochastic procedures preserve the total mass of both species.

\section{Results}
\subsection{Experimental Results}
\subsubsection{Dried Droplet Patterns}
\begin{figure}[pos = htbp]
\centering
  \includegraphics[width=1.0\linewidth]{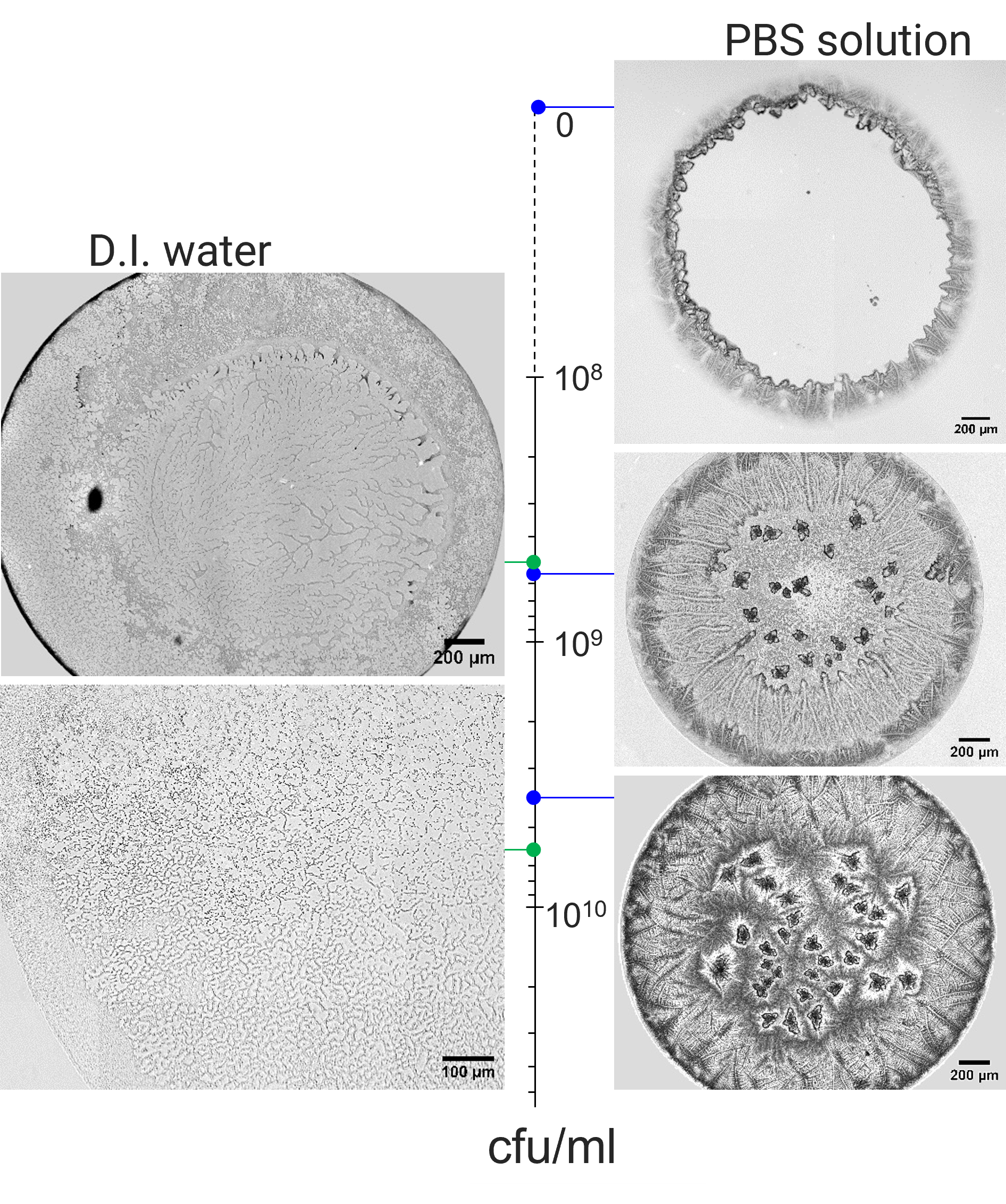}
 \caption{Dried sessile droplet patterns as a function of \textit{E. coli} cell concentration (cfu/ml) in either deionized (D.I.) water or PBS solution. For droplets in D.I. water, the patterns correspond to bacterial deposition, whereas for droplets in PBS solution, they represent salt precipitation patterns, as the bacteria are overshadowed by the precipitated salts.}
  \label{fig:driedpattern}
\end{figure}

\begin{figure}[pos = htpb]
\centering
  \includegraphics[width=\linewidth]{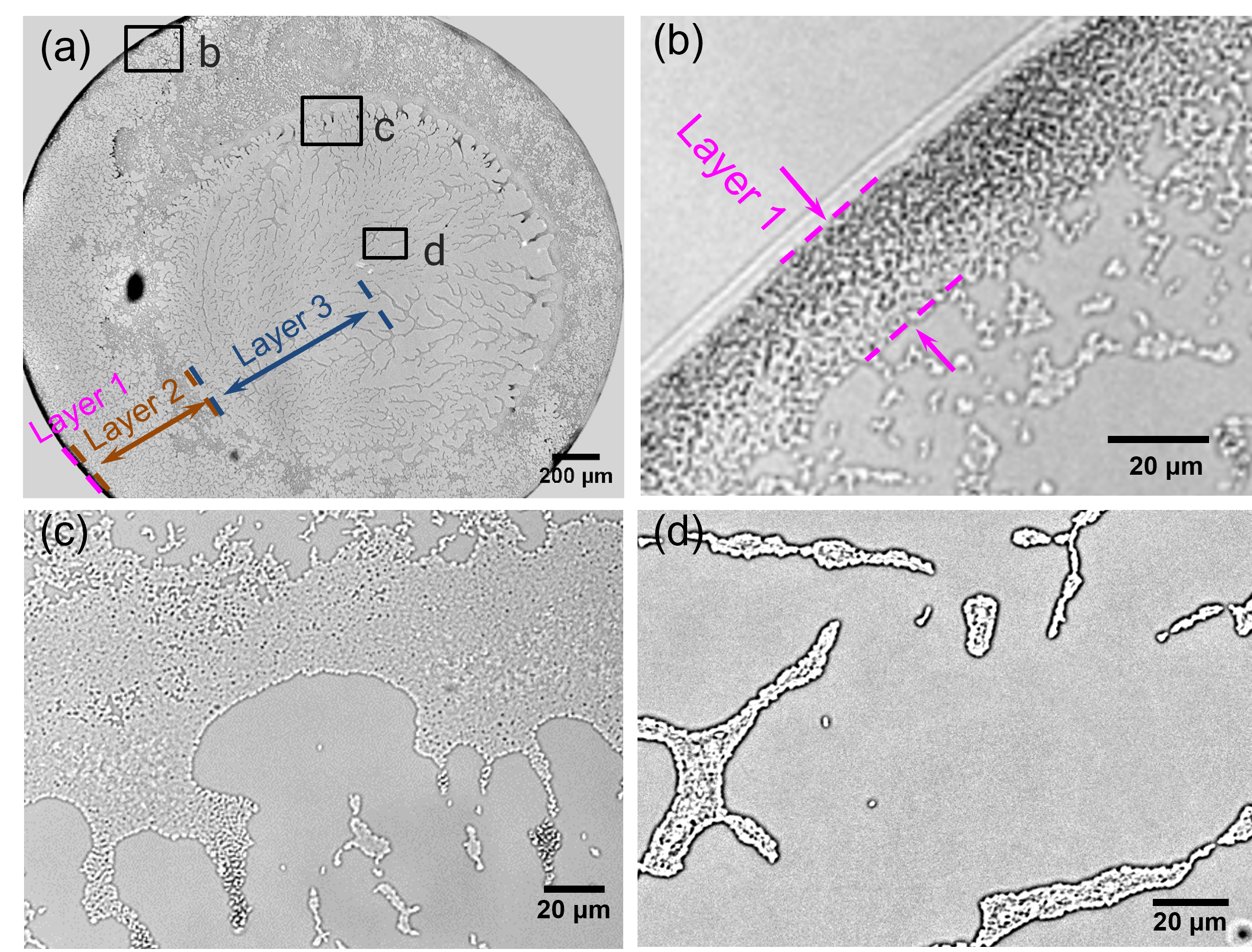}
  \caption{Detailed dried droplet pattern of \textit{E. coli} ($4\times10^8$ cfu/ml) in D.I. water: (a) overall pattern and zoomed in at positions (b), (c) and (d).}
  \label{fig:closelook1}
\end{figure}

Fig.~\ref{fig:driedpattern} summarizes the dried droplet patterns, where the left column corresponds to droplets containing \textit{E. coli} in D.I. water and the right column to those in PBS solution. Images are arranged in the order of \textit{E. coli} concentration (cfu/ml). These tests are repeated multiple times and the dried pattern for each condition remains consistent (see \nameref{Supp_Info} S9). The general trends are summarized below: 

At a low bacterial concentration ($4\times10^8$ cfu/ml) in D.I. water, the dried droplet exhibits a spatially heterogeneous, three-tier deposition pattern. This baseline morphology, also shown in Fig.~\ref{fig:closelook1}(a), reveals a highly concentrated peripheral ring and a progressively dilute, structurally complex interior. Fig.~\ref{fig:closelook1}(b) highlights the outermost peripheral layer (layer 1), which consists of a densely packed, monolayer-thick ring of \textit{E.~coli} spanning approximately $30 \mu m$. More evidence that the ring is monolayer-thick can be found from \nameref{Supp_Info} S3 and S4, which shows the last-second drying video clip and a time series image taken at 1 minute interval throughout the drying process.This dense accumulation indicates that primary transport is dominated by the evaporation-induced outward capillary flow, which sweeps the suspended bacteria toward the pinned triple-phase contact line. The restriction to a single-layer thickness suggests that geometric constraints at the wedged contact line ultimately limit further vertical stacking. Moving radially inward, the intermediate zone (layer 2) (Fig.~\ref{fig:closelook1}(b) and (c)) displays a marked reduction in overall bacterial density, transitioning into a patchy, aggregated deposition. This patchiness likely arises during the later stages of evaporation and we hypothesize three mechanisms contribute to the patch pattern: (i) a substantial amount of \textit{E. coli} is accumulated at the air-water interface over time, whose evidence will be provided in the numerical modeling in Section~\ref{sec:modeling} and Fig.~\ref{fig:con_modeling}. (ii) The bacteria are trapped at this interface that is hard to escape due to the interfacial energy barrier and (iii) the lateral capillary forces (the Cheerios effect) take over to cause the bacteria to clump into patches rather than depositing smoothly~\citep{lewandowski2010orientation, madivala2009self}.  At the boundary between layer 2 and the innermost zone (layer 3) (Fig.~\ref{fig:closelook1}(c)), a distinct fingering instability emerges. This plume morphology likely arises from an asynchronous depinning instability~\citep{deegan2000contact}. While the droplet contact line is largely pinned throughout the entire drying process (evidenced in Fig.~\ref{fig:timesnap} and \nameref{Supp_Info} S4), the contact line one-bacteria-thickness above the substrate is receding and has an uneven stick-slip receding rate due to the non-uniform deposition of the bacteria. Collective motion of motile bacteria may also contribute to this self-organized periodic pattern~\citep{dombrowski2004self,wilting2025active}. Within the central region (layer 3), the macroscopic bacterial concentration is dilute; however, the bacteria self-assemble into complex, dendritic structures (Fig.~\ref{fig:closelook1}(a) and (d)). The stems of these branches are oriented roughly radially outward, aligning with the trajectory of the evaporating fluid flux. While the surrounding area is mostly devoid of cells, the bacteria within these branches remain densely packed. These branched structures are also possible to emerge from an evaporation-induced dewetting instability, where the uneven inward retreat of the thin fluid film coupled with strong interparticle attraction, drives the self-assembly of the remaining suspended cells into extended radial networks flow~\citep{crivoi2013fingering}.

As the bacteria concentration increases to $5\times10^9$ cfu/ml, the deposition morphology transitions into a distinct two-layer structure (Fig.~\ref{fig:driedpattern}). The outermost region forms a thick, highly concentrated peripheral ring spanning approximately 100 $\mu$m in width, separated from the interior by a sharp boundary. The emergence of this distinct interface can be attributed to the temporal evolution of the evaporation-induced flow~\citep{marin2011order}. During the early stages of drying, the outward fluid velocity is relatively moderate, allowing the arriving \textit{E.~coli} sufficient time to rearrange into a dense, ordered packing via Brownian motion and active self-propulsion. However, as the droplet nears complete evaporation, the capillary flow velocity undergoes a temporal singularity and surges dramatically. This rapid, late-stage flux kinetically arrests the remaining cells, abruptly jamming them into a disordered, lower-density phase that physically defines the sharp inner boundary of the peripheral ring. Within the inner second layer, the \textit{E.~coli} assemble into branched clusters that progressively break down into smaller, more segregated branches toward the droplet center. The prominent fingering and plume instabilities observed at the intermediate boundary of the dilute case are entirely suppressed at this higher concentration.

To establish a pure-solute baseline, we examined the evaporation of a droplet containing only the PBS solute (Fig.~\ref{fig:driedpattern}). As expected, the capillary-driven outward flow transports the dissolved salts to the pinned perimeter, resulting in a classic coffee-ring deposit. However, the resulting crystalline ring exhibits a distinct morphological transition. At the outermost contact line, the crystals exhibit a highly branched, dendritic structure that grows inward, which then transitions into larger, blocky bulk crystals at the inner boundary of the ring (particularly on the left-top side of the dried pattern in Fig.~\ref{fig:driedpattern}). This structural shift suggests a transition in crystallization kinetics driven by a strong local supersaturation gradient. The high evaporation rate at the extreme edge induces rapid, diffusion-limited dendritic growth~\citep{nenow1977formation}. As the local solute depletes and the driving force decreases toward the inner boundary, the kinetics shift to slower, interface-limited growth, allowing the formation of faceted bulk crystals~\citep{nenow1977formation}.

\begin{figure}[pos = htpb]
\centering
  \includegraphics[width=\linewidth]{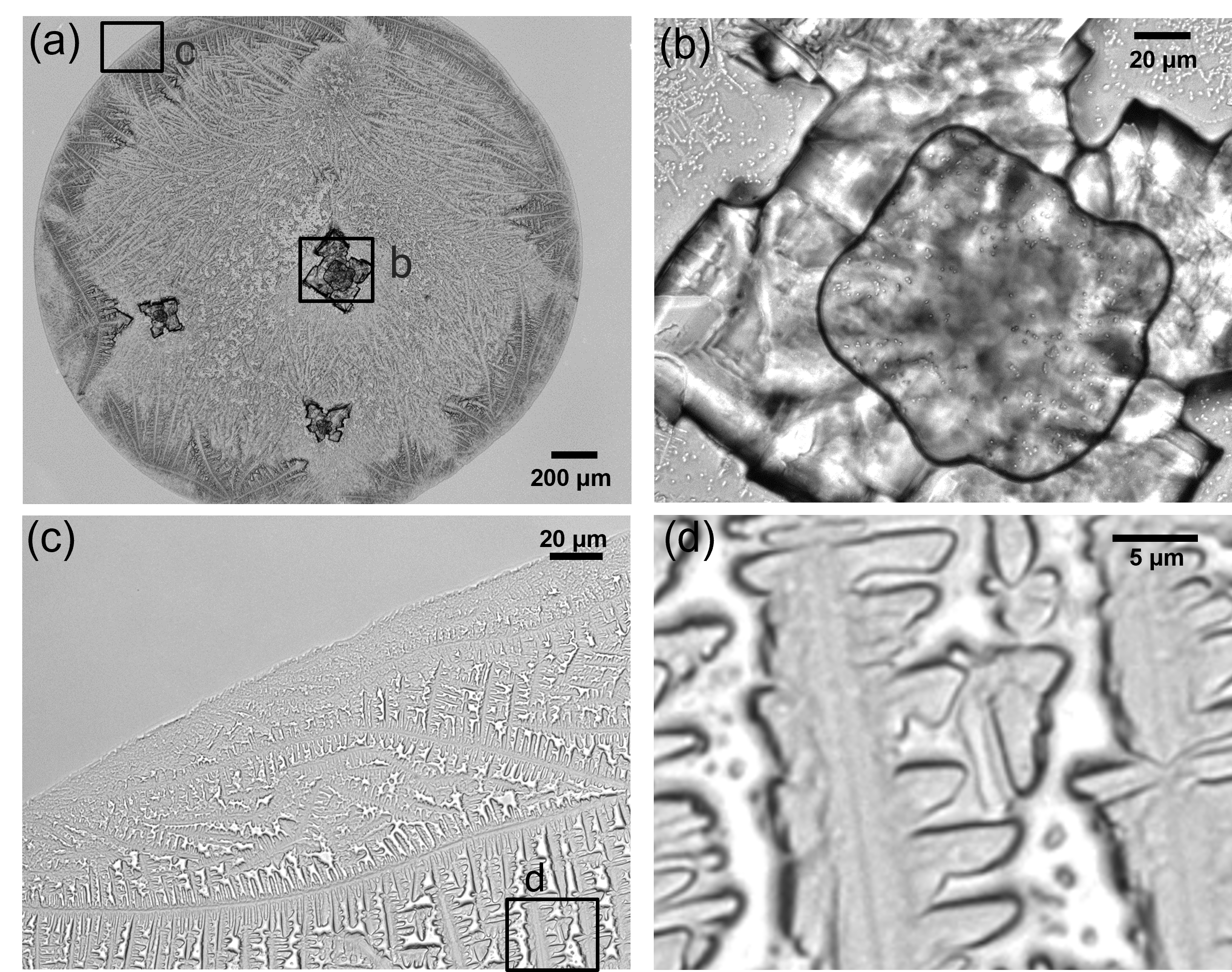}
  \caption{Detailed dried droplet pattern of \textit{E. coli} ($4\times10^8$ cfu/ml) in PBS solution: (a) overall pattern and zoomed in at positions (b), (c) and (d). Deposited \textit{E.~coli} cells can be observed around the crystallized salts or left trace on them.}
  \label{fig:closelook2}
\end{figure}

When \textit{E.~coli} is introduced to the PBS solution at concentrations of $4 \times 10^8$ and $5 \times 10^9$ cfu/ml, the resulting dried patterns demonstrate a profound coupling between the biological phase and salt crystallization (Fig.~\ref{fig:driedpattern}). In both scenarios, robust dendritic crystals emerge from the pinned perimeter and extend radially inward. These outer formations eventually terminate at intermediate radial distances, beyond which the central region is populated by isolated "island" crystals. However, the specific crystalline architecture is highly sensitive to the initial bacterial concentration. At the lower concentration ($4 \times 10^8$ cfu/ml), Fig.~\ref{fig:driedpattern} and the close-up Fig.~\ref{fig:closelook2} show that the central island crystals are sparsely distributed, separated by regions of pure bacterial deposition devoid of localized salt crystallization (see the crystal-free bacteria stains in Fig.~\ref{fig:closelook2}(b) around the central crystal). Furthermore, the inward-growing dendritic branches clearly terminate, transitioning into standard faceted bulk crystal morphologies. Conversely, at the higher bacterial concentration ($5 \times 10^9$ cfu/ml), the entire droplet footprint is covered by crystallized salt. The interior is densely populated with closely neighboring island crystals, and the outer dendritic networks are significantly more compacted, retaining their branched, fractal-like structure without transitioning into single-crystal morphologies.These concentration-dependent discrepancies are driven by the interplay of heterogeneous nucleation and the rheological influence of biological polymers~\citep{kaya2010pattern, alipour2022controlling}. The massive influx of cells at the higher concentration provides an abundance of active nucleation sites across the droplet, driving the high density of interior island crystals.  Similar to the morphological complexities reported in synthetic polymer-salt mixtures~\citep{kaya2010pattern}, salt solute diffusion is likely restricted by the elevated presence of Extracellular Polymeric Substances (EPS) excreted by the bacteria. This environmental constraint suppresses standard interface-limited bulk growth, instead forcing the salt to precipitate into the densely packed, globally distributed dendritic architectures observed at high cell concentrations.

\subsubsection{Drying Process}
\begin{figure*}[pos = t]
\centering
  \includegraphics[width=0.8\textwidth]{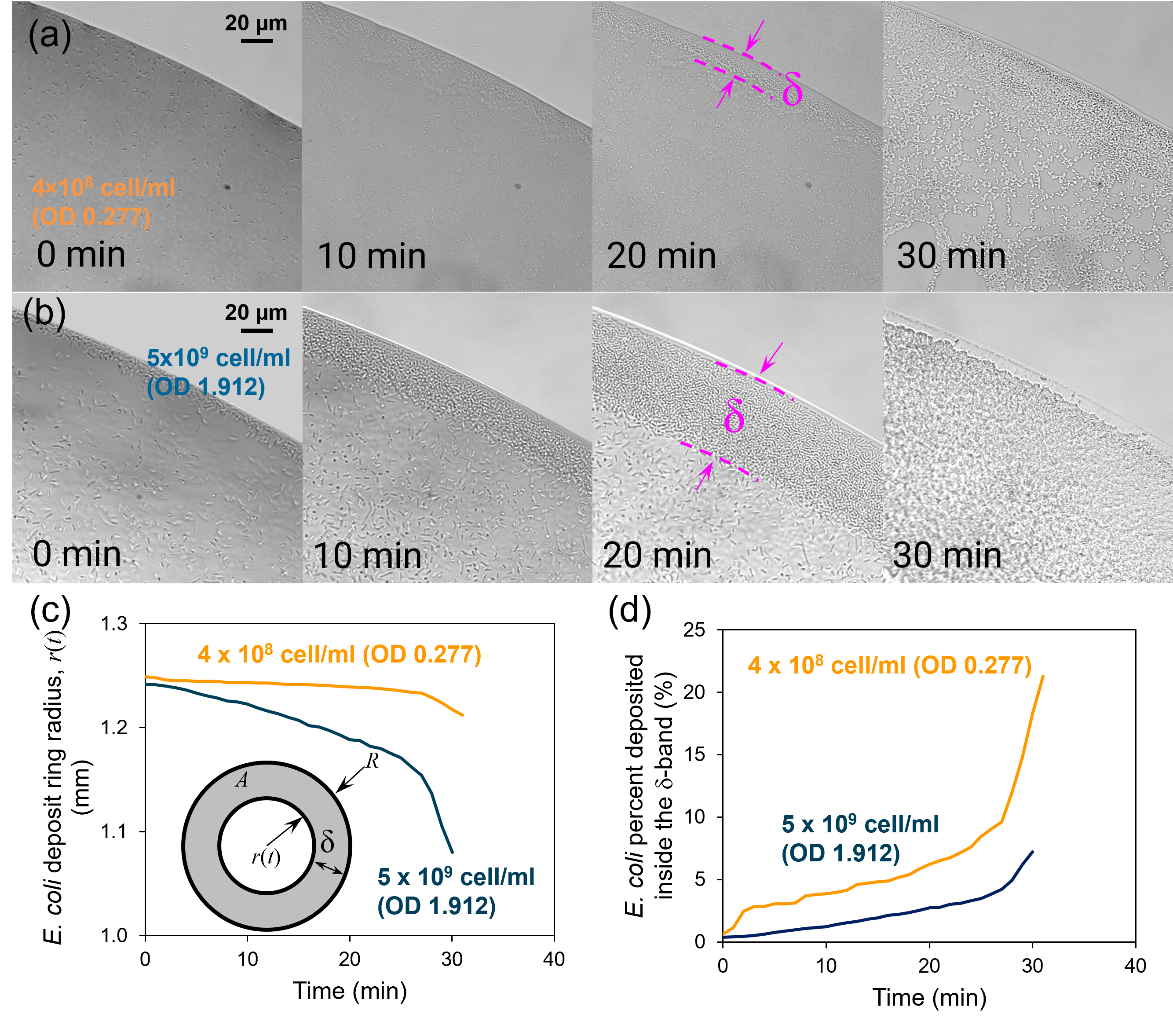}
  \caption{Time-step snapshots of two D.I. water droplets with different \textit{E. coli} concentration: (a) cfu = $4\times10^8$ cfu/ml; (b) cfu = $5\times10^9$ cfu/ml (c) Bacteria deposit ring radius $r(t)$. Note: $ (r(t) = R - \delta(t))$ and (c) Bacteria percent that is deposited inside the $\delta$-band. }
  \label{fig:timesnap}
\end{figure*}

In-situ optical microscopy and time-lapse imaging of the drying process provide critical insights into the dynamic formation process of these patterns. Fig.~\ref{fig:timesnap}(a) and (b) present time-series snapshots of the \textit{E.~coli}-laden droplets in D.I. water (refer to \nameref{Supp_Info} S4 for the real-time dynamics of the dilute case). For both concentrations, bacteria accumulate near the pinned contact line, naturally aligning their longitudinal axes parallel to the boundary. Once swept to this perimeter and deposited, the cells become physically immobilized. In contrast, bacteria within the bulk fluid remain mobile (see \nameref{Supp_Info} S2) until the final stages of evaporation. During this terminal phase, the diverging capillary flux becomes so intense that active bacterial motility is entirely overwhelmed by the convective fluid motion (see \nameref{Supp_Info} S3). A strong recirculating vortex emerges just before complete drying, rapidly sweeping the remaining unattached bacteria toward the substrate. This accelerated, chaotic deposition prevents the cells from rearranging into optimal packing configurations, resulting in a loosely packed and structurally disordered inner region.To quantify these deposition kinetics, the growth of the outer layer was measured by tracking its thickness, $\delta$. As shown in Fig.~\ref{fig:timesnap}(c), $\delta$ increases gradually throughout the evaporation process before exhibiting a sharp, nonlinear surge at the very end, corresponding to the late-stage flux divergence. Interestingly, Fig.~\ref{fig:timesnap}(d) reveals that a significantly higher percentage of the total \textit{E.~coli} population is deposited within this $\delta$-band in the dilute scenario compared to the high-concentration scenario. This discrepancy arises because, at elevated concentrations, severe cell crowding promotes early attachment to the bulk substrate, trapping a larger fraction of the bacteria before they can be mobilized toward the contact line. Both active motility and complex cell-cell hydrodynamic interactions likely contribute to this premature bulk adhesion. Finally, it is important to note that the droplets containing \textit{E.~coli} suspended in the PBS solution follow an nearly identical transport trajectory compared to in D.I.water, with the only notable deviation being the rapid precipitation of salt crystals during the final 1-2 seconds of evaporation.
  
\subsubsection{\textit{E.~coli} Motility Quantification}\label{sec:motility}
\begin{figure}[pos = htbp]
\centering
  \includegraphics[width=\linewidth]{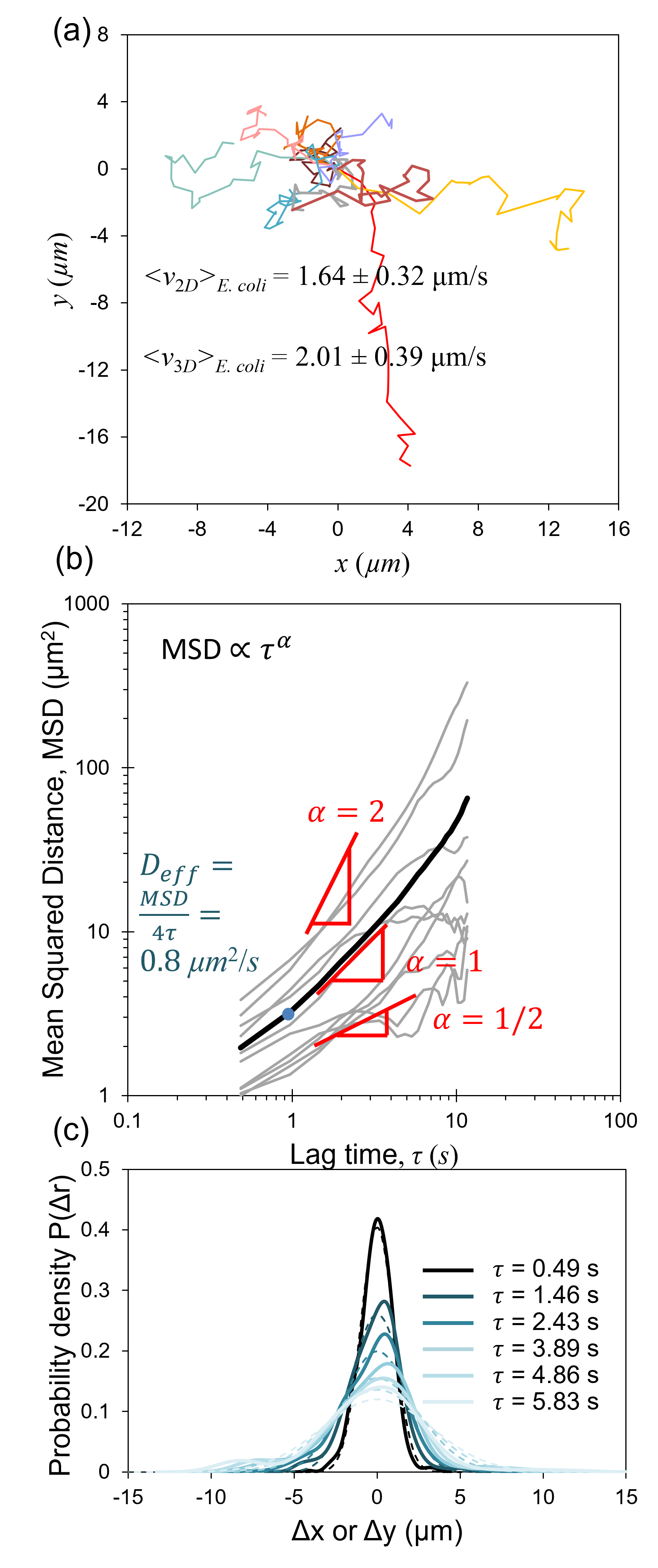}
  \caption{\textit{E. coli} motility analysis: (a) 2$D$ trajectory of 10 randomly picked bacteria over 12.64 s (background drift flow subtracted); (b) Mean Squared Travel Distance (MSD) versus lag time $\tau$ (Black curve: averaged); (c) Probability density $P(\Delta r)$ versus linear travel distance $\Delta x$ or $\Delta y$ for various lag time $\tau$ (continuous curve: experimental data; dashed curve: Gaussian curve with same variance).}
  \label{fig:diffusivity}
\end{figure}

\textit{E.~coli} motility was quantified from a 12.6~s video clip taken $70\,\mu{m}$ above the substrate, 3 minutes after initiation. At this stage, the evaporation-induced background flow was weak (see \nameref{Supp_Info} S2 for raw footage and S1 for further explanation). We randomly selected 10 bacteria to serve as passive tracers and calculated their average velocity, denoted as $\langle \mathbf{v}_{2D} \rangle_{{bg}}$, to represent the background flow. The intrinsic swimming velocity of the bacteria was then obtained by subtracting this background contribution from the raw observed mean velocity:
\begin{equation}
\langle \mathbf{v}_{2D} \rangle_{\textit{E.~coli}} = \langle \mathbf{v}_{2D} \rangle_{{obs}} - \langle \mathbf{v}_{2D} \rangle_{{bg}}
\end{equation}
where $\langle \mathbf{v}_{2D} \rangle_{{obs}}$ is the raw mean velocity directly extracted from the video. Figure~\ref{fig:diffusivity}(a) shows the trajectories of the \textit{E.~coli} tracers over the 12.6~s observation period. This yields a mean 2D velocity of $\langle v_{2D} \rangle_{\textit{E.~coli}} = 1.64 \pm 0.32\,\mu{m}/{s}$, corresponding to a 3D velocity of $\langle v_{3D} \rangle_{\textit{E.~coli}} = 2.01 \pm 0.39\,\mu{m}/{s}$. 

Figure~\ref{fig:diffusivity}(b) displays the Mean Squared Displacement (MSD) versus the lag time $\tau$ for the ten cells, with the bold black curve indicating the averaged relationship. MSD is the average of the squared distances a tracer moves over a given time interval (quantifying the spatial spread of its trajectory), and $\tau$ is the designated time scale. The relationship ${MSD} \propto \tau^{\alpha}$ defines the fundamental mode of motion: $\alpha = 1$ indicates normal Brownian diffusion (random thermal motion), $\alpha > 1$ indicates super-diffusion (active swimming or directed flow), and $\alpha < 1$ indicates sub-diffusion (restricted, confined, or hindered motion). The non-linear $\log$-$\log$ relationship suggests that the background flow is not perfectly offset. Assuming $\alpha = 1$ at a characteristic lag time $\tau = 1\,{s}$, the effective diffusivity of the motile \textit{E.~coli} is $D_{{eff}} = {MSD} / (4\tau) = 0.8\,\mu{m}^2/{s}$. While this is lower than the $30-300\,\mu{m}^2/{s}$ range typically reported for actively motile bacteria~\citep{wilson2011differential,saragosti2012modeling}, it remains higher than the $0.3\,\mu{m}^2/{s}$ value characteristic of non-motile bacteria~\citep{wilson2011differential}. Because the average bacterial motion aligns with $\alpha \approx 1$, the effective diffusivity $D_{{eff}}$ sufficiently characterizes bacterial motility for the subsequent analyses.

Figure~\ref{fig:diffusivity}(c) presents the probability density function $P(\Delta r)$, which represents the likelihood that a tracer moves a specific distance ($\Delta x$ or $\Delta y$) over a lag time $\tau$. The continuous and dashed curves represent the experimental data and the corresponding Gaussian fits, respectively. Across all values of $\tau$, $P(\Delta r)$ peaks near zero, indicating isotropic movement. The experimental curves closely align with the Gaussian fits, exhibiting only marginally higher peaks, which confirms that the bacterial motion is predominantly Brownian.

\subsection{Numerical Modeling}\label{sec:modeling}
\subsubsection{Evaporation-Induced Flow Field}
\begin{figure*}[pos = t]
\centering
  \includegraphics[width=\linewidth]{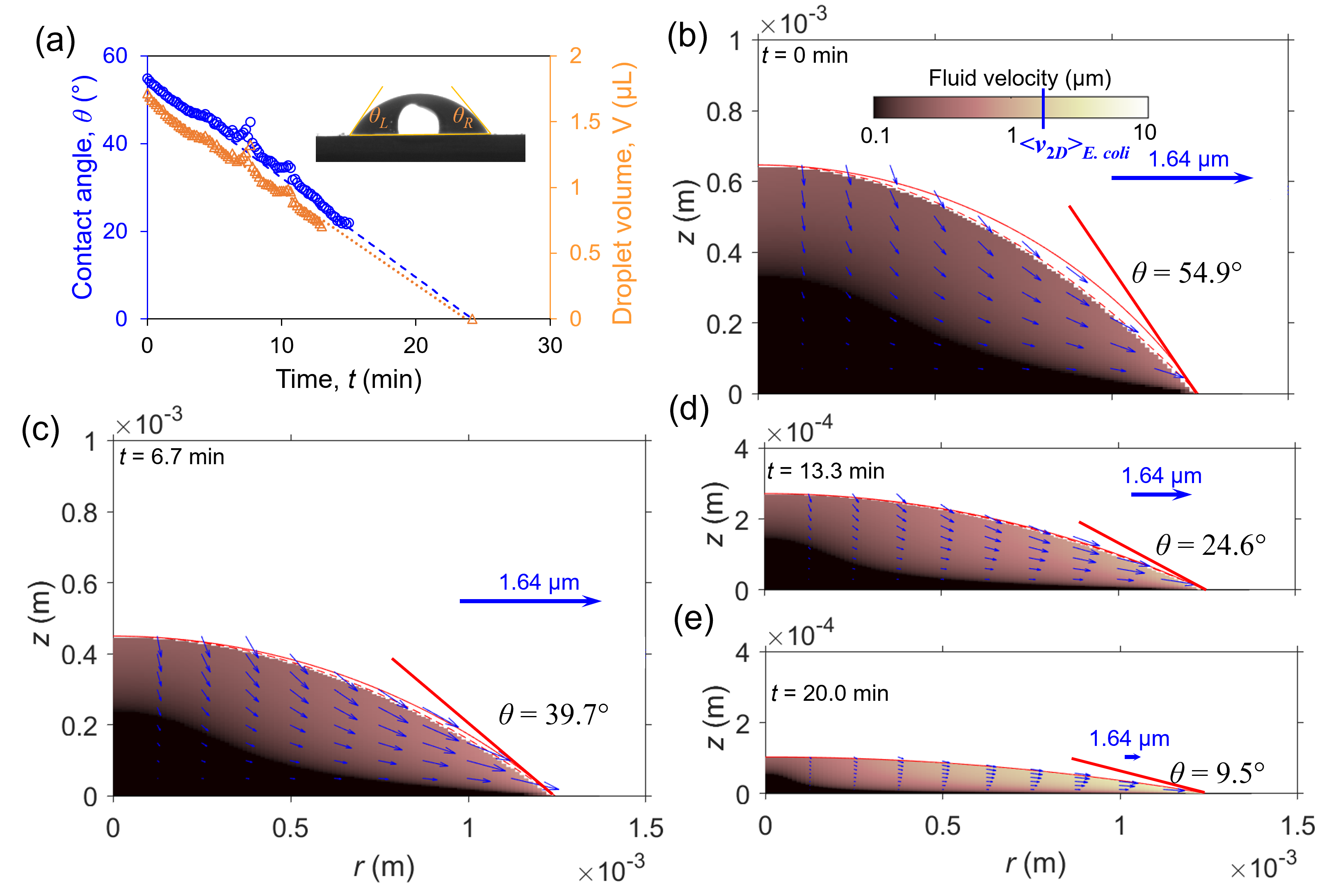}
  \caption{Fluid flow analysis: (a) contact angle and droplet volume evolution with time (insert: photo taken to calculate contact angle and volume); (b)-(e) fluid flow velocity at different stages of evaporation.}
  \label{fig:flowfield}
\end{figure*}

To quantify \textit{E. coli} motility against background fluid flow, and thus the contribution of their motility to the redistribution of the salt precipitation pattern, the velocities of bacteria and fluid flow were determined. Figure~\ref{fig:flowfield} shows the calculated flow field. Figure~\ref{fig:flowfield}(a) indicates that both the contact angle and the droplet volume reduce linearly over time, which provides a necessary characterization of the total evaporation flux for the fluid field calculation. Figure~\ref{fig:flowfield}(b)-(d) show the sequence of evaporation induced flow inside the droplet at time $t =$ 0, 6.7, 13.3 and 20.0 min with the flow field indicated by the blue arrows and the background color indicates the flow magnitude. Note the flow velocity relative to $<v_{2D}>_{E.~coli}$ is also compared at each time snap. The results show that \textit{E.~coli} is able to swim against the background flow and potentially get trapped to the substrate far from the periphery for the majority of the time other than the final stage. This velocity comparison certainly shows that the bacteria motility has the potential to alter the dried deposition patterns.

\subsubsection{Transport and Deposition of \textit{E.~coli} and Salt}

\begin{figure*}[pos = t]
\centering
\includegraphics[width=\linewidth]{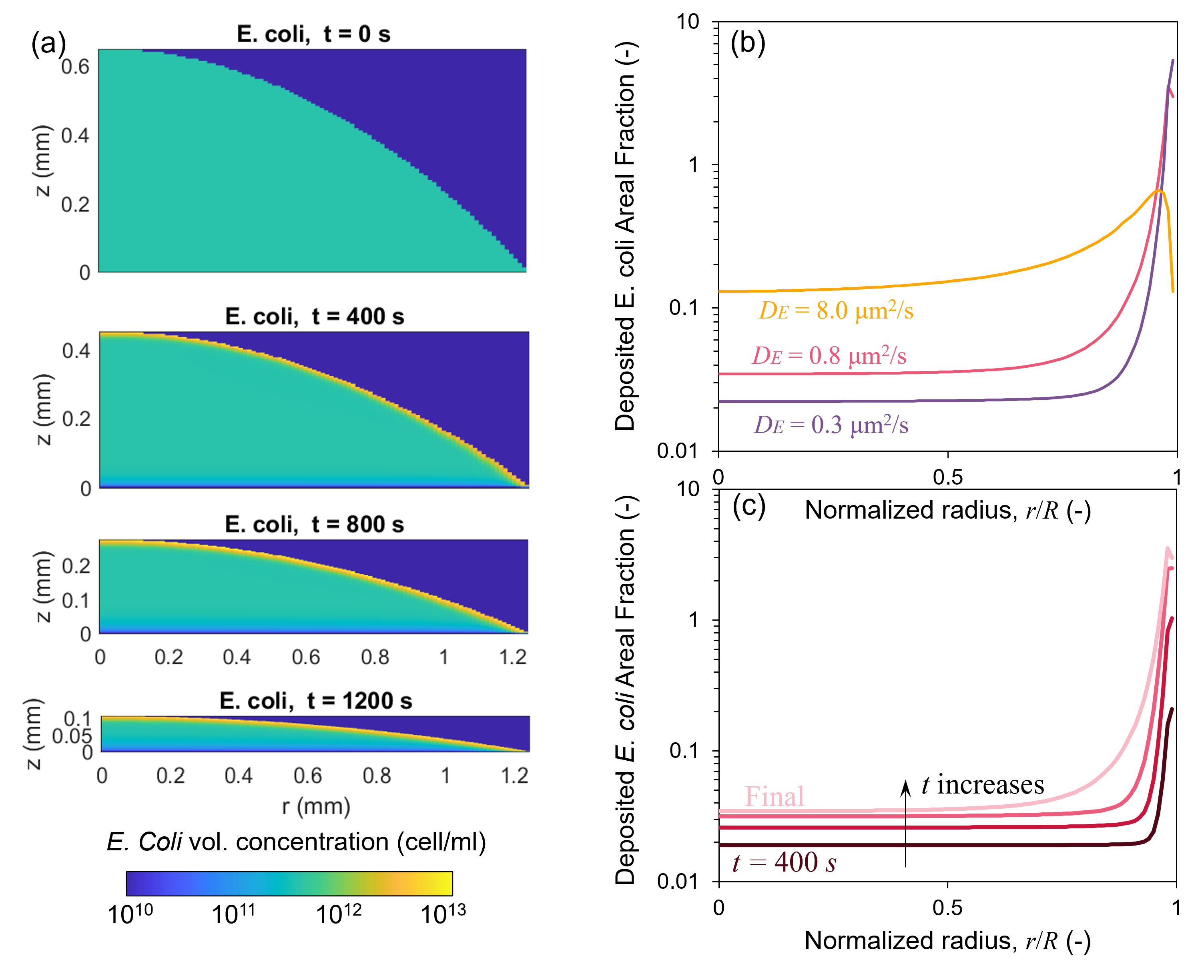}
  \caption{\textit{E. coli} concentration evolution as the droplet dries out: (a) cell concentration in the bulk volume; (b) Cell deposition distribution on the substrate when dried, based on different estimation of $D_E$; (c) Cell deposition distribution on the substrate at $t$ = 400, 800, 1200, and final of 1450s when $D_E = 0.8 \mu m^2/s$. All cases use \textit{E.~coli} concentration ($4\times10^8$ cfu/ml).}
  \label{fig:con_modeling}
\end{figure*}

\begin{figure}[pos = htbp]
\centering
\includegraphics[width=\linewidth]{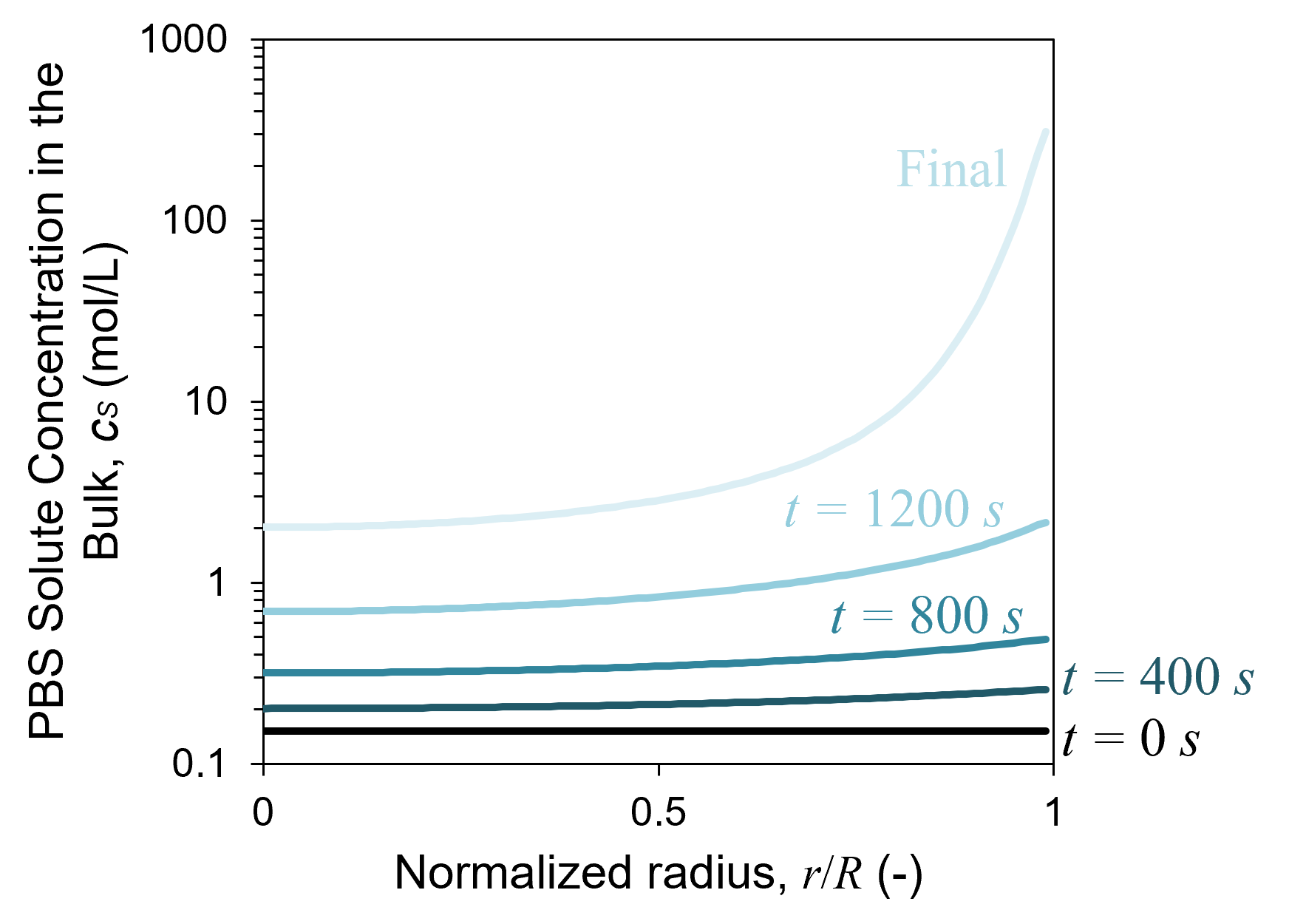}
  \caption{Depth-averaged PBS solute distribution in the bulk over time.}
  \label{fig:salt_con}
\end{figure}

Figure~\ref{fig:con_modeling}(a) displays a series of computed time-lapse snapshots of the \textit{E.~coli} distribution within the bulk fluid. Bacteria tend to accumulate at the free surface due to the outward evaporation flux, while simultaneously depleting near the substrate as a result of the adsorption process. Figure~\ref{fig:con_modeling}(b) demonstrates that compared to non-motile bacteria ($D_E = 0.3 \, \mu m^2/s$), motile bacteria ($D_E = 0.8 \, \mu m^2/s$) tend to accumulate more within the interior of the droplet rather than concentrating heavily at the perimeter. For the highest motility case ($D_E = 8.0 \, \mu m^2/s$), this trend becomes even more pronounced. At such a high diffusivity, the peak concentration is no longer situated at the edge; the bacteria easily overcome the outward background flow, swimming to the substrate at nearly any radial location. Additionally, the greater droplet height near the center allows more bacteria to attach there compared to the shallower edges. These modeling results support our hypothesis that bacterial motility can fundamentally alter and even reverse typical deposition patterns. Figure~\ref{fig:con_modeling}(c) illustrates the evolution of deposited bacteria over time for the baseline case ($D_E = 0.8 \, \mu m^2/s$). Bacteria tend to accumulate at both the center and the edge, but the deposition rate at the edge increases more rapidly due to the stronger outward fluid flux. Figure~\ref{fig:salt_con} shows the evolution of the depth-averaged salt concentration over time, which increases monotonically and remains comparatively uniform in the radial direction until the very late stage of evaporation.

\subsubsection{Salt Crystallization Patterns}
\begin{figure*}[pos = t]
\centering
\includegraphics[width=\linewidth]{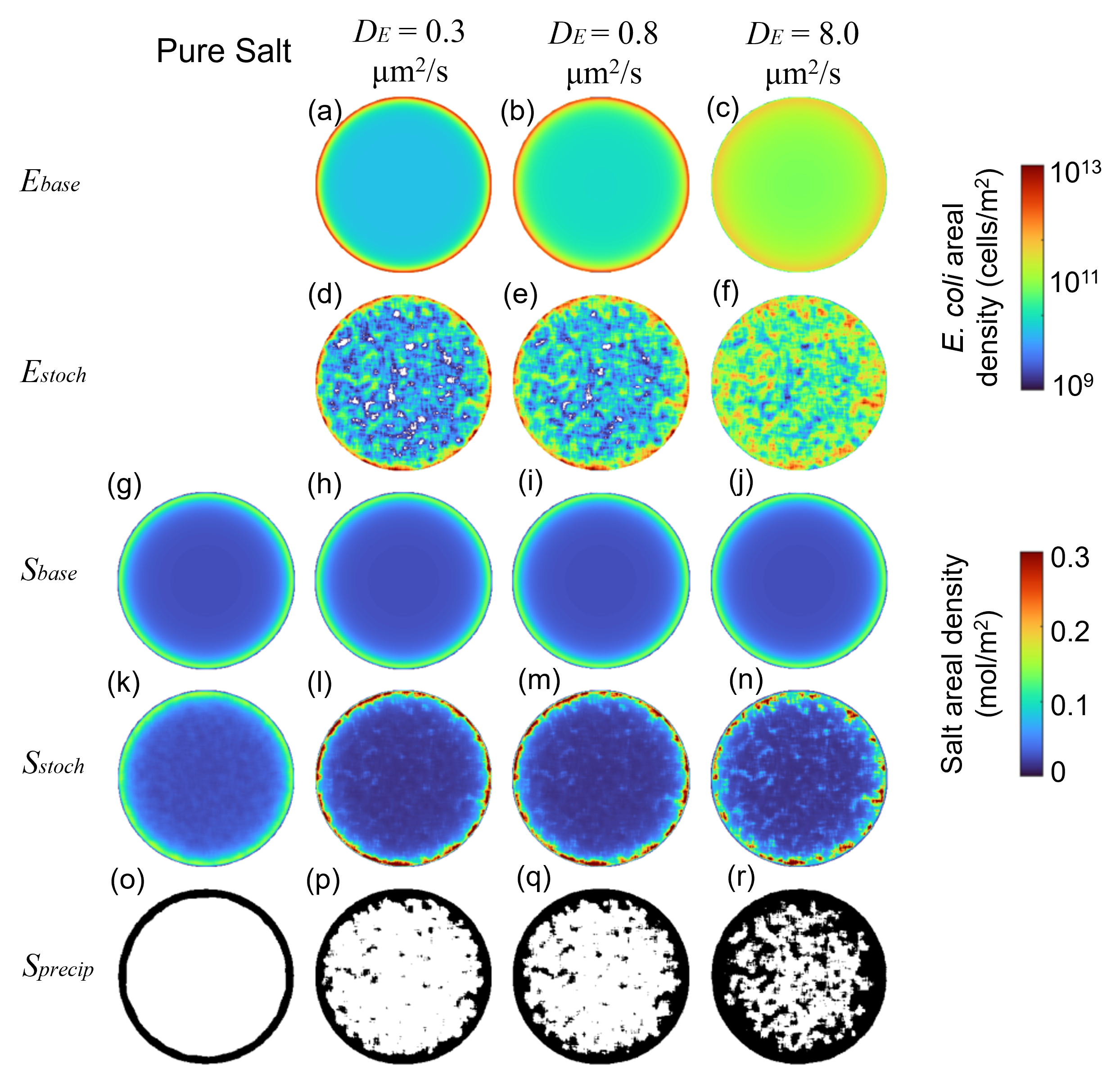}
  \caption{Simulated deposition and precipitation patterns. Four columns represents a pure salt case and three \textit{E~.coli}-presented cases when $D_E = 0.3, 0.8, 8.0 \mu m^2/s$. Row 1 (a)(b)(c): Baseline \textit{E.~coli} areal density $E_{base}$ on the substrate mapped from the 1D case; row 2 (d)(e)(f): \textit{E.~coli} areal density after 2D stochastic treatment; Row 3 (g)(h)(i)(j): baseline salt areal density mapped from 1D, which is the same for the four cases; Row 4 (k)(l)(m)(n): salt areal density after the stochastic treatments; row 5 (o)(p)(q)(r): final precipitated salt patterns (black: crystallized area; white: no crystallization).All cases use an initial bulk \textit{E.~coli} concentration ($4\times10^8$ cfu/ml).}
  \label{fig:pattern_modeling}
\end{figure*}

As shown in Fig.~\ref{fig:pattern_modeling}(a)-(c), the calculated 1D \textit{E.~coli} substrate concentrations are mapped into the 2D axisymmetric baseline fields $E_{base}$ for three representative bacteria diffusivities $D_E= 0.3, 0.8, 8.0 \mu m^2/s$, while Fig.~\ref{fig:pattern_modeling}(g)-(j) depicts the corresponding 2D salt deposition patterns, $S_{base}$, which are identical across the four cases. The stochastic fields $E_{stoch}$ and $S_{stoch}$ obtained after the randomized field layer is superimposed are illustrated in Fig.~\ref{fig:pattern_modeling}(d)-(f) for \textit{E~.coli} and Fig.~\ref{fig:pattern_modeling}(k)-(n) for salt. As shown in Fig.~\ref{fig:pattern_modeling}, these stochastic perturbations introduce realistic spatial variation without obscuring the underlying background concentration gradients.

The resulting salt crystallization patterns are shown in Fig.~\ref{fig:pattern_modeling}(o)-(r) and exhibit strong qualitative agreement with experimental observations (Fig.~\ref{fig:driedpattern}). In the pure salt scenario, crystallization occurs exclusively along the droplet periphery. For the moderate-motility case, $D_E = 0.8 \mu m^2/s$, salt continues to precipitate predominantly at the outer boundary, accompanied by isolated central patches. Non-motile \textit{E.~coli} ($D_E = 0.3 \mu m^2/s$) yield fewer isolated central crystal sites, whereas highly motile bacteria ($D_E = 8.0 \mu m^2/s$) promote substantial central salt precipitation, forming interconnected crystal domains in several regions.

Overall, the simulation results demonstrate that enhanced bacterial motility—modeled as an elevated isotropic diffusivity $D_E$—is a key governing factor in dictating salt crystallization patterns of an evaporated sessile droplets.

\section{Discussions}
\subsection{Insight into Active Particle Altered Precipitation}

Our experiments show that motile \textit{E.~coli} in an evaporating salt-solute droplet can alter the resulting crystallization pattern, and our numerical model qualitatively supports the hypothesis that this alteration is driven by bacterial motility, which creates locally favorable nucleation sites. This suggests that salt crystallization patterns could be deliberately engineered by controlling active particles such as motile bacteria or other active nano- to macro-particles, rather than treating deposition patterns as a passive byproduct of transport alone.

Unlike passive tracers, active particles self-propel and can be guided, either externally or through their own directed responses to environmental gradients (chemotaxis, phototaxis, thermotaxis), to accumulate in specific regions of an evaporating droplet. This offers a mechanism for spatiotemporal, programmable control over precipitation patterning that passive-particle systems cannot achieve. Active motility can also enhance local mixing, producing super-diffusive transport and altering the macroscopic suspension viscosity~\citep{sokolov2009reduction,jeong2026rheology}, which opens rich physical insights and new avenues for engineering control.

Our current model does not resolve the crystal morphology observed in our experiments, but active particles may also have an impact on this. Active particles can act as physical impurities that preferentially adsorb onto specific crystal faces, a well-documented habit-modification mechanism in crystal growth~\citep{boistelle1982impurity}, biasing anisotropic face-growth rates and reshaping the resulting crystal habit. Motility-driven local flow may also perturb the diffusive boundary layer surrounding a growing nucleus, shifting the balance between diffusion-limited and kinetics-limited growth that governs whether crystals develop as smooth, faceted forms or as branched, dendritic structures~\citep{mullins1963morphological}. Finally, active particles may become physically trapped within the advancing crystal lattice as fluid inclusions, as has been directly observed for bacteria entrapped during laboratory halite growth~\citep{adamski2006entrapment}, introducing defects, inclusions, or grain boundaries that further alter crystal texture. Disentangling these morphology-altering pathways, beyond the nucleation-site framework treated here, is a natural direction for future work.

\subsection{Limitations}

The findings above are subject to a number of experimental and modeling limitations, which are collected here. 

In the experiments, the bacterial population used in this study was not the most motile achievable. The measured mean three-dimensional swimming velocity is 6.8 times slower than the value reported for the same strain~\citep{jeong2026rheology} and falls below the $5-40\,\mu{m}/{s}$ range typical of motile bacteria~\citep{wilson2011differential}; likewise, the extracted effective diffusivity $D_{{eff}} = 0.8\,\mu{m}^2/{s}$ lies well below the $30-300\,\mu{m}^2/{s}$ range reported for actively motile cells. This is attributed to the absence of a motility buffer and to metabolic decay over the course of the experiment, so the observed population is a mixture of motile and no-longer-motile cells. 
 
In the computational analysis, the coupling between the flow and the transported species is considered one-way. Marangoni and thermal-gradient effects are not included, which could exist in the experiments, and is a known mechanism for recirculation that could redistribute cells and ions in the droplet interior. In the species transport model, the two species are transported independently, so the salt deposition field $S_{base}$ is identical across all cases and the electrostatic attraction of dissolved ions to the negatively charged cell envelope is not resolved dynamically but collapsed into a single redistribution step immediately before dryout; this ties the predicted degree of co-localization to a phenomenological coupling rather than to a transport calculation. Representing motility as an isotropic effective diffusivity discards run-and-tumble kinematics, cell orientation and rod shape, chemotaxis, cell-cell hydrodynamic interactions, and collective motion, and therefore cannot reproduce any directed or self-organized accumulation, only an enhanced spreading.
 
In the stochastic crystallization model, the model predicts where salt precipitates, not how it grows, thus, it does not resolve crystal morphology. The stochastic parameters, including the baseline nucleation threshold $S_0$, the Gaussian kernel widths, the noise amplitudes, and the neighbor-growth bias, are phenomenological and were chosen to reproduce the qualitative trends rather than calibrated against independent measurements. Consistent with this, the modeling framework as a whole is intended to demonstrate qualitatively that bacterial motility is one of the leading factors altering the salt crystallization pattern, not to replicate the experimentally observed physics quantitatively.
 
\section{Conclusions}

We show experimentally that introducing motile \textit{E.~coli} into a drying PBS droplet transforms the classic coffee-ring salt deposit into a two-part pattern: peripheral crystals that grow increasingly dendritic and radially extended, and isolated interior "island" crystals whose number and density scale with bacterial concentration. Neither feature appears in the pure-salt control.

Combined Stokes-flow, finite-volume transport, and stochastic nucleation model confirms both of our original hypotheses: (1) Bacteria act as heterogeneous nucleation sites that locally lower the salt supersaturation threshold, and (2) bacterial motility is sufficient to let cells swim against the outward capillary flow during early evaporation and populate the droplet interior rather than being swept entirely to the contact line. Numerical simulation also suggests that varying only this single motility parameter across three representative values (characterized by the effective diffusivity $D_E = 0.3, 0.8, 8.0\ \mu$m$^2$/s) can result in different salt crystallization patterns: nucleation confined to the periphery when bacteria are absent or immobile, a handful of isolated interior crystals at moderate motility, and dense, interconnected interior crystal domains as motility increases.

These results reframe evaporative salt crystallization from a passive, transport-limited process into one that active particles can dynamically steer. The finding that motility itself, rather than mere bacterial presence, serves as a controllable lever over deposition patterns points toward new strategies for engineering evaporative crystallization. Such a study has future applications in various engineering technologies involving biosensing, printed electronics and others.

\printcredits

\newcounter{suppitem}
\newcommand{\suppitem}[1]{%
    \refstepcounter{suppitem}%
    \item S\thesuppitem\_#1 \label{Supp_S\thesuppitem}%
}

\section*{Supplementary Information}\label{Supp_Info}
Additional information related to this work can be found in the following 9 supplementary files (\url{https://sandy.unl.edu/record/139?&ln=en}):
\begin{itemize}
    \suppitem{Supplementary\_Information\_Summary.docx}
    \suppitem{Bacteria\_in\_the\_bulk.avi}
    \suppitem{Dried\_moment.avi}
    \suppitem{Interface\_Time\_Series\_Dilute\_Ecoli\_Water.gif}
    \suppitem{Sessile\_Drop\_Analytical\_ReplicateHuLarson2005.m}
    \suppitem{Sessile\_Drop\_Analytical.m}
    \suppitem{SoluteTransport.m}
    \suppitem{StochasticCrystalization.m}
    \suppitem{Repeated\_Dried\_Patterns.docx}
\end{itemize}

\section*{Declaration of competing interests}
The authors declare that they have no known competing financial interests or personal relationships that could have appeared to influence the work reported in this paper.

\section*{Declaration of Generative AI Use}
The authors used Claude Opus 4.8 and DeepSeek-V3 to help implement the authentic numerical modeling ideas into code. After that, the code was carefully reviewed and revised, and the authors take the full responsibility of the code.

\section*{Acknowledgments}
This material is based upon work supported by the National Science Foundation (CMMI-1943722). Any opinions, findings and conclusions, or recommendations expressed in
this material are those of the authors and do not necessarily reflect those of the NSF. This work was performed at Optical Microscopy Core at Georgia Tech Petit Biotechnology Building through the Shared User Management System (SUMS). We also thank for the startup funding from the University of Nebraska - Lincoln. 

\section*{Data availability}
The Appendix and the Supplementary Information provide additional information that is sufficient to reproduce the results. More data can be obtained upon reasonable request to the corresponding author. 

\appendix
\label{appendix}

\renewcommand{\thefigure}{\thesection\arabic{figure}}
\renewcommand{\theHfigure}{appendix.\thesection.\arabic{figure}}

\section{Appendix}
\setcounter{figure}{0}
\setcounter{equation}{0}
\renewcommand{\theequation}{A\arabic{equation}}
\renewcommand{\thefigure}{A\arabic{figure}}
We present here a summary of the theories used to compute the evaporation induced flow field, the transport and deposition of \textit{E. coli} and salt based on the flow field, as well as the salt crystallization morphology based on the final \textit{E.~coli} and salt deposition pattern. 

\subsection{Replication of the Flow Field From Hu \& Larson, 2005~\citep{hu2005analysis}}\label{App_Hu}
The analytical model for the flow field induced by evaporation in the sessile droplet follows the work by Hu \& Larson, 2005~\citep{hu2005analysis}. We first replicated the analytical solution from Hu \& Larson, 2005~\citep{hu2005analysis} in the MATLAB code (see Supplementary Information S\ref{Supp_S1} and S\ref{Supp_S5}). A brief summary of the equation used follows.

The evaporation mass flux $J$ [ Mass / (Time $\cdot$ Length$^2$] at the free surface is a function of axisymmetric coordinate $r$ and time $t$ given the pinned constant droplet radius $R$:
\begin{equation}
J(r,t) = J_0(\theta) \left( 1 - \left(\frac{r}{R}\right)^2 \right)^{-\lambda(\theta)}
\end{equation}

where
\begin{align}
J_0(\theta, t) = &\frac{D c_v (1 - H)}{R} (0.27\theta^2 + 1.30) \nonumber \\
&\left( 0.6381 - 0.2239 \left(\theta - \frac{\pi}{4}\right)^2 \right)\label{eq_J0}
\end{align}

in which $J_0(\theta, t)$ represents the evaporation flux at the center of the droplet ($r = 0$) as a function of the instantaneous contact angle $\theta$. $D$ is the diffusion coefficient of water vapor in air, $c_v$ is the saturated water vapor concentration, and $H$ is the relative humidity of the ambient air. An exponent $\lambda(\theta)$ is introduced to represent the singularity that dictates the power-law divergence of the evaporation flux near the pinned contact line ($r \rightarrow R$), varying with the instantaneous contact angle $\theta$

\begin{equation}
\lambda(\theta) = \frac{1}{2} - \frac{\theta}{\pi}
\end{equation}

Based on our experimental measurement of the sessile droplet contact angle $\theta$ using the ram\'{e}-hart tensiometer (see Figure~\ref{fig:flowfield}(a)), $\theta$ can be approximated as linearly reducing from an initial angle of $\theta_0$ until totally evaporated at time $t_f$

\begin{equation}
\theta = -\theta_0\frac{t}{t_f} + \theta_0
\end{equation}

Then, a dimensionless or scaled average evaporation parameter $\bar{J}(t)$ can be defined, which relates the localized mass loss to the global thinning rate of the droplet:
\begin{equation}
\bar{J}(t) = \frac{-\dot{h}_0(\theta)}{\rho h}
\end{equation}

Where $\dot{h}_0$ is the rate of change of the sessile droplet apex height $h_0$. $\rho$ is the liquid density, and $h = h(r, t)$ is the local height of the droplet profile.  For a spherical cap, the initial apex height $h_0$ is directly related to the initial contact angle $\theta_0$ and radius $R$ as:

\begin{equation}
h_0 = \frac{R}{\sin\theta_0} \left( 1 - \frac{1}{\cos\theta_0} \right)
\end{equation}

The parabolic droplet profile approximation adopted by \citep{hu2005analysis} takes the form:
\begin{equation}
h_{p}(r,t) = h(0, t)\left(1-\left(\frac{r}{R}\right)^2\right)\label{eq:parabolic}
\end{equation}

Finally, the radial and vertical flow velocity $v_r$ and $v_z$ on the cylindrical coordinate $(r, z)$ are obtained by scaling the dimensionless velocity components $\bar{v}_r$ and $\bar{v}_z$ by characteristic geometric lengths ($R$ and $h_0$) and the total evaporation time $t_f$:
\begin{equation}
v_r = \bar{v}_r(t, r, z) R / t_f \label{eq_ur}
\end{equation}
\begin{equation}
v_z = \bar{v}_z(t, r, z) h_0 / t_f \label{eq_uz}
\end{equation}

Where the dimensionless radial velocity component $\bar{v}_r$, capturing the fluid being replenished outward toward the pinned edge due to lubrication flow theory, is:
\begin{align}
& \bar{v}_r =  
\frac{3}{8} \frac{1}{1 - t/t_f} \frac{1}{r/R} \left[ \left(1 - (r/R)^2\right) - \left(1 - (r/R)^2\right)^{-\lambda(\theta)}\right] \nonumber \\
&\Bigg( \frac{z^2}{h_{p}^2} - 2\frac{z}{h_{p}} \Bigg) + \Bigg( \frac{(r/R)h_0^2(h_{p}/h_0)}{R^2} \bigg( \bar{J}\lambda(\theta) \nonumber \\
&\Bigg(1 - \left(\frac{r}{R}\right)^2\Bigg)^{-\lambda(\theta)-1} + 1 \Bigg) \left( \frac{z}{h_{p}} - \frac{3}{2}\frac{z^2}{h_{p}^2} \right)\Bigg)\label{eq_ur_bar}
\end{align}

And the corresponding dimensionless vertical velocity component $\bar{v}_z$, derived from the continuity equation to ensure mass conservation throughout the droplet depth, is:

\begin{align}
&\bar{v}_z =  
\frac{3}{4} \frac{1}{1 - t/t_f}\left( 1 + \lambda(\theta)(1 - (r/R)^2)^{-\lambda(\theta)-1} \right) \nonumber \\
&\left( \frac{z^3/h_0}{3h^2} - \frac{z^2/h_0}{h} \right) + \frac{3}{2} \frac{1}{1 - t/t_f} \nonumber \\
&\left[ (1 - (r/R)^2) - (1 - (r/R)^2)^{-\lambda(\theta)} \right] \left( \frac{z^2}{3h_{p}^2} - \frac{z^3}{3h_{p}^3} \right)  \nonumber \\
& \frac{h(0, t/t_f)}{h_0} - \Biggl( \frac{h_0^2}{R^2} \left( \bar{J}\lambda(\theta)(1 - (r/R)^2)^{-\lambda(\theta)-1} + 1 \right) \nonumber \\
& \left( \frac{z^2}{h_0^2} - \frac{z^3/h_0^3}{h/h_0} \right) + \frac{rh_0^2}{R^3}\bar{J}\lambda(\theta)\left( \lambda(\theta) \right. + 1\biggr) \nonumber \\
&(1 - (r/R)^2)^{-\lambda(\theta)-2} \left( \frac{z^2}{h_0^2} - \frac{z^3/h_0^3}{h_{p}/h_0} \right) - \frac{rh_0^2}{R^3}\bar{J}\lambda(\theta) \nonumber \\
&\left( (1 - (r/R)^2)^{-\lambda(\theta)-1} \right. + 1\Biggr) \left. \left( \frac{z^3/h_0^3}{h_{p}/h_0} \right) \frac{h(0, t/t_f)}{h_0} \right)\label{eq_uz_bar}
\end{align}

By applying the equations, we successfully replicated the analytical and Finite Element Method (FEM) computed flow field by Hu \& Larson, 2005~\citep{hu2005analysis}. 

\subsection{Calculation of the Flow Field in Our Experiments}\label{App_Flow}

After validating the flow field calculation against the scenario reported in the literature~\citep{hu2005analysis}, we implemented the same equations to replicate the flow field in our experimental setting. We also checked that the calculated flow field is divergence-free, which guarantees the conservation of mass of the solute species. From experimental measurement, the initial contact angle was set to $\theta_0$ = 54.9\textdegree, and the total evaporation time $t_f$ = 1450 s. Contact line radius $R$ = 1.245 $mm$, vapor diffusivity $D$ = $1.9\times10^{-5} m^2/s$, and vapor density at the interface $c_v$ = $1.73\times10^{-2}kg/m^3$, corresponding to a 100\% humidity at 21\textdegree C, relative humidity at far space $H$ = 0.4 and droplet density $\rho$ = 1000 $kg/m^3$. See Supplementary Information S\ref{Supp_S1} and S\ref{Supp_S6} for details. 

\subsection{Static Interface Motion in Transformed Coordinate}\label{App_ALE}
Before computing species transport in an evaporating sessile droplet driven by the velocity field derived in Appendix~\ref{App_Hu}, we performed a coordinate transformation and demonstrate here that the droplet free surface becomes static under the new coordinates. This is a typical methodology to transfer an Eulerian framework to an Arbitrary Lagrangian–Eulerian (ALE) framework that eases the computation~\citep{souli2000ale}. For a spherical cap droplet:

\begin{equation}
h(r, t) = \sqrt{\left(\frac{R^2 + h(0, t)^2}{2h(0, t)}\right)^2 - r^2} - \frac{R^2 - h(0, t)^2}{2h(0, t)}
\end{equation}
one has to transfer both the radial and elevation coordinate $(r, z)$ in order to have a static interface, which other studies usually adopt~\citep{souli2000ale,yang2014fully}. On the other hand, we decide to use the parabolic profile approximation following Eq.~\ref{eq:parabolic} and introduce a dimensionless coordinates $(\chi, \xi)$ defined by:
\begin{equation}
\chi = \frac{r}{R}, \qquad \xi = \frac{h}{h(0, t)}
\end{equation}

where $\chi \in [0,1]$ is the normalized radial coordinate and $\xi \in [0,1]$ is the normalized vertical coordinate scaled by the instantaneous apex height. 

\begin{figure}[htbp]
\centering
  \includegraphics[width=0.5\textwidth]{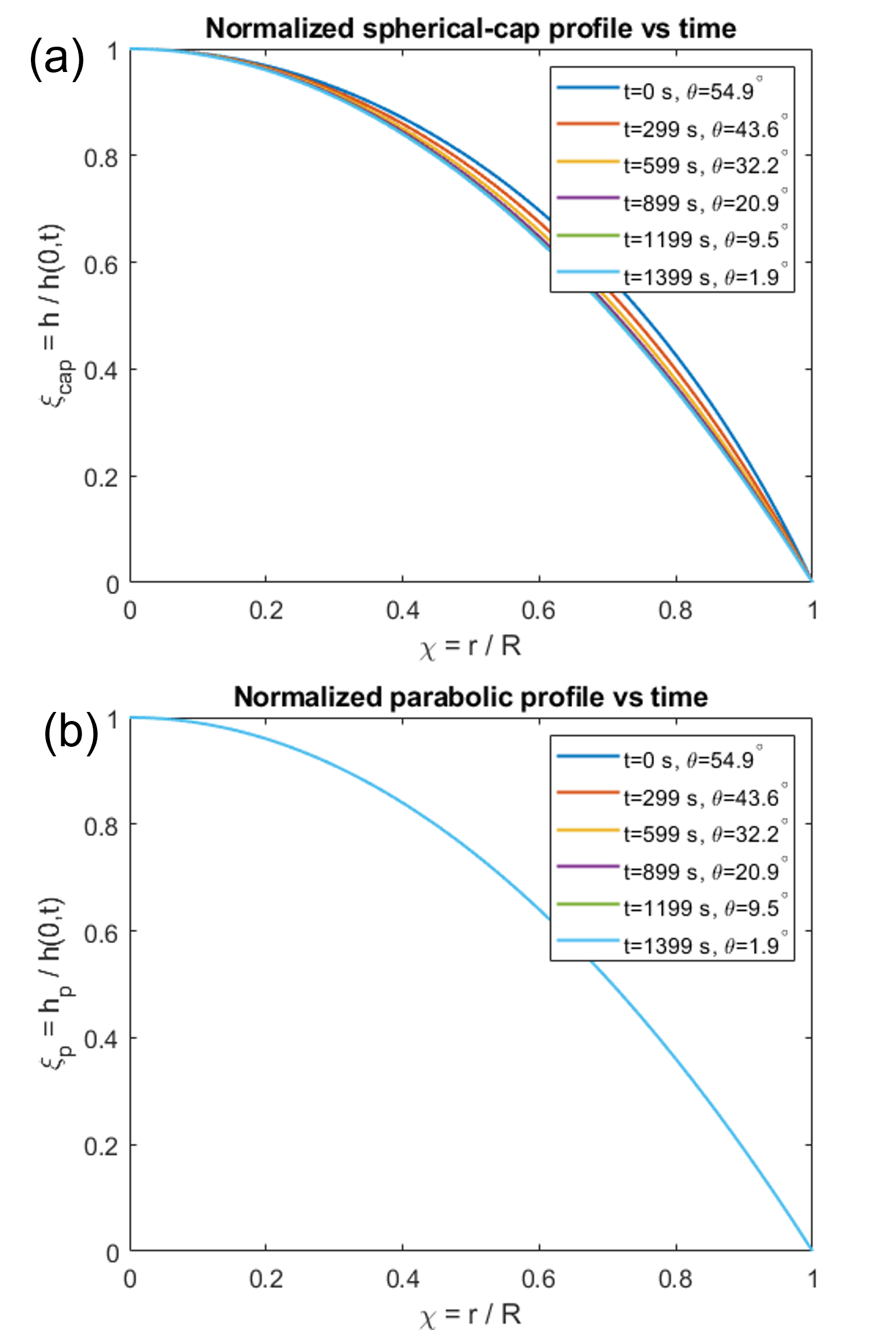}
  \caption{Droplet profile over time on transformed coordinate $(\chi, \xi)$: (a) Moving interface for a spherical cap profile; (b) Static interface for a parabolic approximated profile}
  \label{fig:Appex_ALE}
\end{figure}

Substituting $r = \chi R$ into the spherical-cap profile and normalizing by $h(0,t)$
gives:
\begin{align}
& \xi_{cap}(\chi, t) =   \frac{h}{h(0,t)} = \nonumber \\ &\frac{1}{h(0,t)}  \left[\sqrt{\left(\frac{R^2+h(0,t)^2}{2h(0,t)}\right)^2 - \chi^2 R^2} - \frac{R^2-h(0,t)^2}{2h(0,t)}\right]
\end{align}
which retains an explicit dependence on $h(0, t)$
and therefore continues to evolve in $(\chi, \xi)$
space as the droplet evaporates (See Fig.~\ref{fig:Appex_ALE}(a)). In contrast, substituting $r = \chi R$
into the parabolic approximation and normalizing by $h(0,t)$:
\begin{equation}
\xi_{p}(\chi) = \frac{h_{p}}{h(0,t)} = \frac{h(0,t)\left(1 - \chi^2\right)}{h(0,t)} = 1 - \chi^2
\end{equation}
The apex height $h(0, t)$ cancels exactly, yielding a profile that depends only on $\chi$ with no time dependence (See Fig.~\ref{fig:Appex_ALE}(b)). The free surface is therefore represented by the fixed parabola $\xi = 1 - \chi^2$ throughout the entire evaporation process, regardless of the instantaneous contact angle, until the droplet is totally dried out. We adopt this $(\chi, \xi)$ coordinates for the computational domain in the following sections as it simplifies the in-house solver implementation, particularly suitable for subsequent steps involving species transport and deposition under various boundary conditions.

\subsection{Computing \textit{E. coli} and salt transport using Finite Volume Method}\label{app_Transport}
After the background flow field and droplet profile have been derived from the previous section, this section elaborates the theory adopted to compute \textit{E. coli} and PBS salt solute transport in the bulk and adsorption at the substrate, which produces the Figs.~\ref{fig:con_modeling} and~\ref{fig:salt_con} in the manuscript section~\ref{sec:modeling}. See the code that implements the idea of this section in the Supplementary Information S\ref{Supp_S1} and S\ref{Supp_S7}.

\subsubsection{Transport Equation in Transformed Framework~--~\textit{E.~coli}}

In physical cylindrical coordinates $(r,z,t)$, the \textit{E.~coli} concentration 
$c_{E}$ satisfies the incompressible advection--diffusion equation:
\begin{align}
    & \frac{\partial c_{E}}{\partial t}
    + \frac{1}{r}\frac{\partial(r u_r c_{E})}{\partial r}
    + \frac{\partial(u_z c_{E})}{\partial z}
    = \nonumber \\
    &\frac{1}{r}\frac{\partial}{\partial r}
      \!\left(r D_{E}\frac{\partial c_{E}}{\partial r}\right)
    + \frac{\partial}{\partial z}
      \!\left(D_{E}\frac{\partial c_{E}}{\partial z}\right)
    \label{eq:transport_phys}
\end{align}
where $(u_r, u_z)$ is the velocity field computed through Eqs.~\ref{eq_ur} 
to~\ref{eq_uz_bar}, and $D_{E} = 0.8\mu m^2/s$ is the 
effective diffusivity of \textit{E.~coli} in fluid.

\paragraph{Boundary conditions in $(r,z,t)$:}
At the axis of symmetry ($r=0$), zero radial solute flux is required:
\begin{equation}
    \frac{\partial c_{E}}{\partial r}\bigg|_{r=0} = 0
    \label{eq:axis_BC_rzt}
\end{equation}

At the free surface ($z = h(r,t)$), \textit{E.~coli} cells do not evaporate, 
so the total flux of bacteria through the interface is zero:
\begin{equation}
    c_{{E}} (\bm u - \bm{u}_{{surf}})\cdot \bm {n}
    - D_{E}\,\nabla c_{E}\cdot \bm{n}
    \big|_{z=h_p(r, t)} = 0
    \label{eq:freesurface_BC_total}
\end{equation}
where $\bm{u}_{{surf}}$ is the velocity of the free surface and 
$\bm{n}$ is the outward unit normal.  

At the substrate ($z=0$), we assume \textit{E.~coli} adsorbs irreversibly 
following a linear irreversable (first-order) relationship: 
\begin{equation}
    -D_{E}\frac{\partial c_{E}}{\partial z}
    \bigg|_{z=0}
    = k_a^{E}\,c_{E}\big|_{z=0}
    \label{eq:substrate_BC_rzt}
\end{equation}
where $k_a^{E}$ is the first-order adsorption 
rate constant. This is implemented as a surface flux boundary condition rather 
than a volumetric sink.

We transform Eq.~\eqref{eq:transport_phys} to the static mapped coordinates 
$(\chi, \xi)$, where $\chi = r/R$ and 
$\xi = z/h_0(t)$, with $h_0(t) \equiv h(0,t)$ the instantaneous apex height. 
Under this mapping, the transport equation becomes:
\begin{equation}
\begin{aligned} & \frac{\partial c_{{E}}}{\partial t} + \frac{1}{R \chi} \frac{\partial}{\partial \chi} \Big( \chi \, u_r \, c_{{E}} \Big) + \frac{1}{h_0} \frac{\partial}{\partial \xi} \Big( (u_z - \xi \dot{h}_0) \, c_{{E}} \Big) + \frac{\dot{h}_0}{h_0} c_{{E}} \\ &= \frac{D_{E}}{R^2 \chi} \frac{\partial}{\partial \chi} \left( \chi \frac{\partial c_{{E}}}{\partial \chi} \right) + \frac{D_{E}}{h_0^2} \frac{\partial^2 c_{{E}}}{\partial \xi^2} \end{aligned} \label{eq:transport_mapped_final}
\end{equation}
where $c_{{E}}, u_r$ and $u_z$ are all functions of $(\chi, \xi)$, 
and $\dot{h}_0$ is the rate of change of the apex height.

\paragraph{Boundary conditions in $(\chi,\xi,t)$:}
Under the coordinate transformation, the axis condition becomes:
\begin{equation}
    \frac{\partial c_{E}}{\partial \chi}\bigg|_{\chi=0} = 0
    \label{eq:axis_BC_chi}
\end{equation}
and the substrate condition becomes:
\begin{equation}
    -\frac{D_{E}}{h_0}\frac{\partial c_{E}}{\partial \xi}
    \bigg|_{\xi=0} = k_a^{E}\,c_{E}\big|_{\xi=0}
    \label{eq:substrate_BC_chi}
\end{equation}
For the free surface $\xi_s(\chi)$, the zero total flux boundary condition becomes:
\begin{align}
   & c_{{E}} \left(
        \frac{2\chi h_0}{R} u_r
        + u_z - \dot{h}_0 \xi_s
    \right)
    - \nonumber \\
    &D_{{E}} \left(
        \frac{2\chi h_0}{R^2}\frac{\partial c_{{E}}}{\partial \chi}
        + \frac{1}{h_0}\frac{\partial c_{{E}}}{\partial \xi}
    \right)
    = 0
    \quad \text{at } \xi = \xi_s(\chi)
    \label{eq:ale_zero_flux}
\end{align}

Note that the zero-total-flux conditions, Eq.~\ref{eq:ale_zero_flux} is 
enforced naturally in the FVM discretization. The flux through 
the top face ($\xi = \xi_s(\chi)$) is not computed because there 
is no wet cell above it. This directly imposes a zero-flux condition at 
the free surface. The advective flux associated with the moving interface is automatically 
accounted for by the time-dependent cell volume:
\begin{equation}
    V_{jk}^{wet} = r_j\,h_0(t)\,\Delta r\,\Delta\xi
    \cdot\mathbf{1}_{wet}
\end{equation}
which shrinks as $h_0(t)$ decreases. $\mathbf{1}[\cdot]$ means that the function returns "1" if it is a wet cell and 0 otherwise.

Finally, the initial condition is a spatially uniform \textit{E.~.coli} concentration 
throughout the droplet:
\begin{equation}
    \bar{c}_{E}(\chi, 0) = c_{{E},0}
    \label{eq:cell_IC}
\end{equation}

The substrate adsorption condition Eq.~\eqref{eq:substrate_BC_chi} 
is applied as an explicit surface flux, removing mass from the 
substrate-adjacent layer ($\xi=0$) at each substep and accumulating it 
into the deposited surface mass $\sigma_{{E}}(\chi,t)$:
\begin{equation}
    \frac{d\sigma_{{E},j}}{dt} 
    = k_a^{E}\,c_{{E},j,1}\
      \chi_j\,\Delta \chi
    \label{eq:deposit_update}
\end{equation}
where $c_{{E},j,1}$ is the concentration in the first 
($\xi=0$) layer of column $j$.

\paragraph{Discrete FVM formulation.}
To prevent blow-up of the concentration field as the droplet thins near 
dryout, we track the cell solute mass $q_{jk} = c_{jk} V_{jk}^{wet}$ 
as the conserved variable. The update equation for each wet cell is:
\begin{equation}
    \frac{dq_{jk}}{dt} = -\sum_{\text{faces}} \mathcal{F}_{face},
\end{equation}
where face fluxes are evaluated only between two wet cells. 

The radial flux across the $+r$ face between columns $j$ and $j+1$ is
\begin{equation}
    \mathcal{F}_{r,j+1/2,k} = 
    r_{j+1/2} h_0 \Delta\xi \left(
    \bar{u}_r \, c_{{upw}}
    - D \frac{c_{j+1,k} - c_{j,k}}{\Delta r}
    \right)
\end{equation}
where $\bar{u}_r = \tfrac{1}{2}(u_{r,j,k}+u_{r,j+1,k})$ is the face-centered 
radial velocity and $c_{{upw}}$ is the upwind concentration based on 
the sign of the volumetric flux $r_{j+1/2} h_0 \bar{u}_r \Delta\xi$.

The vertical flux across the $+\xi$ face between layers $k$ and $k+1$ is
\begin{equation}
    \mathcal{F}_{\xi,j,k+1/2} = 
    r_j^{({E})} h_0 \Delta r \left(
    \bar{w}_{{eff}} \, c_{{up}}
    - \frac{D}{h_0^2} \frac{c_{j,k+1} - c_{j,k}}{\Delta\xi}
    \right)
\end{equation}
where $\bar{w}_{{eff}} = \frac{1}{2}(w_{{eff},k} + w_{{eff},k+1})$ 
and $w_{{eff}} = u_z/h_0 - \xi \dot{h}_0/h_0$ is the grid-relative 
vertical velocity.

The discrete mass balance for cell $(j,k)$ is then:
\begin{equation}
    \frac{dq_{jk}}{dt} = 
    \left(\mathcal{F}_{r,j-1/2,k} - \mathcal{F}_{r,j+1/2,k}\right)
    + \left(\mathcal{F}_{\xi,j,k-1/2} - \mathcal{F}_{\xi,j,k+1/2}\right)
\end{equation}
with the convention that fluxes through dry faces are zero. The mass is 
updated explicitly:
\begin{equation}
    q_{jk}^{n+1} = q_{jk}^n + \Delta t \left(\frac{dq_{jk}}{dt}\right)^n.
\end{equation}

The timestep is controlled by the combined advective--diffusive CFL condition:
\begin{equation}
    \Delta t = \frac{{\mathrm{CFL}}}{\dfrac{u_{\max}}{\Delta r} + \dfrac{w_{\max}}{\Delta\xi} + 2D_{E}\left(\dfrac{1}{\Delta r^2} + \dfrac{1}{h_0^2 \Delta\xi^2}\right)}
    \label{eq:cfl_ecoli}
\end{equation}
with safety factor $\mathrm{CFL} = 0.01$. This ensures numerical stability 
throughout the evaporation process.

\subsubsection{Transport Equation in Transformed Framework~--~Salt 
(Depth-Averaged)}
Although the PBS solution has more than one salt component, we simplify it as a single salt species with total compound molarity $c_S = 0.15 mol/L$ and use $D_S = 
10^3~\mu m^2/s$. Since an explicit FVM is used, the large salt diffusivity would impose a prohibitively small timestep due to the vertical diffusion stability limit. We therefore adopt a depth‑averaged formulation for the salt concentration. In order to justify the depth-averaging approach, we estimated the P\'{e}clet number along $z$ direction, defined as the ratio of the vertical advection to diffusion timescale: 
\begin{equation}
    {Pe}_z^{S} = \frac{U_z\,h_0}{D_S}
    \approx \frac{h_0^2}{D_S\,t_f}
    \label{eq:Pe_z}
\end{equation}
where the characteristic vertical velocity is estimated as $U_z \sim 
h_0/t_f$ from the kinematic recession of the free surface. 
Using $h_0 \sim 6\times10^{-4}$~m at $t=0$, one obtains:
\begin{equation}
    {Pe}_z^{S}   \sim 0.1\text{--}1
    \label{eq:Pe_z_values}
\end{equation}
For salt, ${Pe}_z^{S} \leq 1$ throughout evaporation, therefore, the salt concentration $c_S$ is rapidly homogenized across the droplet depth and may be treated as vertically uniform:
\begin{equation}
    c_{S}(\chi,\xi,t) \approx \bar{c}_{S}(\chi,t)
    \label{eq:salt_wellmixed}
\end{equation}
This well-mixed assumption allows us to collapse the 
two-dimensional salt problem to a one-dimensional radial equation via 
depth-averaging. By contrast, for \textit{E.~coli}, ${Pe}_z^{E} \gg 1$ 
and Eq.~\eqref{eq:salt_wellmixed} does not hold: strong vertical 
concentration gradients persist throughout evaporation and the full 
two-dimensional treatment described in the preceding section must be retained.

We define the depth-averaged salt concentration as:
\begin{equation}
    \bar{c}_{S}(\chi,t)
    = \frac{1}{\xi_s(\chi)}\int_0^{\xi_s(\chi)} c_{S}(\chi,\xi,t)\,d\xi
    \label{eq:salt_da_vars}
\end{equation}
where $\xi_s  = 1-\chi^2,$ is the static normalized wet-column height (the parabolic free surface in mapped coordinates).  
The corresponding column-integrated salt mass per radian, stored at radial cell center $\chi_j$, is:
\begin{equation}
    Q_j(t) = \chi_j \, h_0(t) \, \xi_s(\chi_j) \, \bar{c}_{S}(\chi_j,t) \, \Delta\chi \, R
    \label{eq:Q_def}
\end{equation}
where $\Delta\chi$ is the uniform grid spacing in the mapped coordinate, 
and $\chi_j = (\chi_{j+1/2}^2-\chi_{j-1/2}^2)/(2\Delta\chi)$ 
is the area‑weighted cell‑center radius.  
Starting from the mapped transport equation for a passive scalar, 
Eq.~\eqref{eq:transport_mapped_final}, with \(c_{{E}}\) replaced by 
\(c_{{S}}\), and invoking the vertical well‑mixed assumption 
\(\partial_\xi c_{{S}} = 0\), we integrate over the local film 
height \(\xi \in [0, \xi_s(\chi)]\). This yields the vertically integrated 
equation:
\begin{equation}
\begin{aligned}
    \xi_s(\chi)\frac{\partial \bar{c}_{S}}{\partial t}
    &+ \frac{1}{R\chi}\frac{\partial}{\partial \chi}
      \!\left(\chi\,\xi_s(\chi)\,\bar{u}_r\,\bar{c}_{S}\right)
      + \frac{\bar{c}_{S}}{h_0}
        \bigl(u_z|_{\xi_s}-\xi_s(\chi) \dot{h}_0\bigr) \\
    &+ \frac{\dot{h}_0}{h_0} \xi_s(\chi) \bar{c}_{S}
    = \frac{D_{S} \xi_s(\chi)}{R^2\chi}
      \frac{\partial}{\partial \chi}
      \left(\chi \frac{\partial \bar{c}_{S}}{\partial \chi}\right)
    \label{eq:salt_da_intermediate}
\end{aligned}
\end{equation}
where the depth‑averaged radial velocity is defined as
\begin{equation}
    \bar{u}_r(\chi,t) = \frac{1}{\xi_s(\chi)}
    \int_0^{\xi_s(\chi)} u_r(\chi,\xi,t)\,d\xi
    \label{eq:ubar_def}
\end{equation}

One can prove that the surface velocity and evaporative flux terms (3rd and 4th terms to the \textit{L.H.S.} of the Eq.~\ref{eq:salt_da_intermediate}) cancel exactly.  
This cancellation is a direct consequence of the moving ALE frame and the 
incompressibility constraint. Defining the \textit{integrated radial velocity}
\[
\tilde{u}_r(\chi,t) = \xi_s(\chi)\,\bar{u}_r(\chi,t)
= \int_0^{\xi_s(\chi)} u_r(\chi,\xi,t)\,d\xi ,
\]
the final \textit{source-free} conservative equation for the column mass is:
\begin{align}
    &\frac{\partial}{\partial t}
    \Bigl( h_0(t)\,\xi_s(\chi)\,\bar{c}_{S} \Bigr)
    + \frac{1}{R\chi}\frac{\partial}{\partial \chi}
    \Bigl( \chi\,h_0(t)\,\tilde{u}_r(\chi,t)\,\bar{c}_{S} \Bigr) \nonumber\\
    &=
    \frac{D_{S}}{R^2\chi}
    \frac{\partial}{\partial \chi}
    \left(
    \chi\,h_0(t)\,\xi_s(\chi)\,
    \frac{\partial \bar{c}_{S}}{\partial \chi}
    \right)
\label{eq:salt_da_final}
\end{align}

Equation~\eqref{eq:salt_da_final} is discretized using an explicit upwind scheme FVM 
on the radially staggered grid with the update
\begin{equation}
Q_j^{n+1} = Q_j^n + \Delta t (F_{j-1/2} - F_{j+1/2}) 
\end{equation}
where the column mass $Q$ is stored at cell centers following Eq.~\ref{eq:Q_def}, and 
the total flux through the right face of cell \(j\) is:
\begin{align}
    &F_{j+1/2} = r_{j+1/2} h_0 \nonumber \\
    &\left(
    \tilde{u}_r^{j+1/2} \bar{c}_{S, j+1/2}^{{upw}}
    - D_{S} \xi_s^{j+1/2}
    \frac{\bar{c}_{S, j+1} - \bar{c}_{S, j}}{\Delta r}
    \right)
\end{align}
 
\paragraph{Boundary and initial conditions:}
Because the salt is non‑volatile and does not adsorb onto the substrate, 
no flux crosses the substrate or the free surface after depth‑averaging. 
The only remaining spatial direction is radial.

At the axis of symmetry (\(\chi=0\)), axisymmetry requires:
\begin{equation}
    \left.\frac{\partial \bar{c}_{S}}{\partial \chi}
    \right|_{\chi=0} = 0 
    \label{eq:salt_BC_axis}
\end{equation}
In our code Supplementary Information S\ref{Supp_S7}, there is not an explicit line to implement this boundary condition and that is because in the FVM discretization, no flux is computed at the leftmost face $(j = 1, \chi = \chi_{1/2} = 0)$, and thus the axisymmetry condition is naturally enforced.

At the triple-contact point (\(\chi=1\)), the local film height vanishes 
\((\xi_s(1)=0)\), so the diffusive flux in Eq.~\eqref{eq:salt_da_final} 
is identically zero. Numerically, we impose:
\begin{equation}
    \left.\frac{\partial \bar{c}_{S}}{\partial \chi}
    \right|_{\chi=1} = 0
    \label{eq:salt_BC_CL}
\end{equation}
which is consistent with setting the outer-face diffusive flux to zero in the 
FVM scheme. Therefore, the code naturally imposes this boundary condition without an explicit coding line as well.

Finally, the initial condition is a spatially uniform salt concentration 
throughout the droplet:
\begin{equation}
    \bar{c}_{S}(\chi, 0) = c_{{S},0}
    \label{eq:salt_IC}
\end{equation}

\subsection{Stochastic Approach for \textit{E. coli} and Salt Deposition Pattern}\label{app_crystal}

After obtaining radially averaged transport outputs, we construct two-dimensional stochastic maps on the circular substrate domain \(\Omega\) (Supplementary Information S\ref{Supp_S8}). The workflow has two stages: (i) stochastic mapping of 1D radial fields to 2D areal-density fields; (ii) nucleation-limited precipitation mapping with mass conservation.

\subsubsection{Stochastic reconstruction of 2D \textit{E. coli} and salt fields}

Let $r \in [0,R]$ denote the radial coordinate, and let $\bm{x} = (x,y) \in \Omega \subset \mathbb{R}^2$ denote a point on the substrate, with radial distance $|\bm{x}|$. Let $\tilde{e}(r)$ and $\tilde{s}(r)$ denote the final one-dimensional radial areal-density profiles used to reconstruct the two-dimensional \textit{E. coli} and salt fields, respectively. 

We define the 
radial areal-density profiles used for the 2D reconstruction as the 
terminal-time values of the column-integrated quantities from that 
section, normalized by the annular cell area 
$A_j = \tfrac{1}{2}\bigl(r_{j+1/2}^2 - r_{j-1/2}^2\bigr)$:
\begin{equation}
    \tilde{e}(r_j) \;=\; \frac{\sigma_{{E},j}(t_f)}{A_j}
    \qquad
    \tilde{s}(r_j) \;=\; \frac{Q_j(t_f)}{A_j}
    \label{eq:ebar_sbar_def}
\end{equation}
Here $\sigma_{{E},j}(t_f)$ is the deposited (substrate-adsorbed) 
\textit{E.~coli} mass per radian defined in 
Eq.~\eqref{eq:deposit_update}, and $Q_j(t_f)$ is the column-integrated 
salt mass per radian defined in Eq.~\eqref{eq:Q_def}, both at the final stage $t_f$, and are 
per-radian quantities, so the factor of $2\pi$ cancels in the ratio 
and $A_j$ needs not be multiplied by an angular extent. 

The corresponding base two-dimensional fields are defined by radial mapping:
\begin{equation}
E_{{base}}(\bm {x}) = \tilde{e}(|\bm{x}|)
\qquad
S_{{base}}(\bm{x}) = \tilde{s}(|\bm{x}|)
\end{equation}

To introduce spatial heterogeneity, Gaussian random fields are generated independently for \textit{E. coli} and salt. Let $\mathcal{N}_E(\bm{x})$ and $\mathcal{N}_S(\bm{x})$ denote zero-mean unit-variance Gaussian random fields. These fields are patch-smoothed by convolution with a local averaging kernel:
\begin{equation}
K(\bm{x}) = \frac{1}{P^2}\mathbf{1}_{P\times P}\label{eq:kernel}
\end{equation}
where $P$ is the patch size chosen here as $1/25R = 50 \mu m$ and $\mathbf{1}_{P\times P}$ denotes a $P\times P$ matrix of ones. The smoothed fields are then normalized by their maximum absolute magnitude, yielding
\begin{equation}
\widehat{\mathcal{N}}_E(\bm{x}) = \frac{(\mathcal{N}_E * K)(\bm{x})}{\displaystyle \max_{\bm{y}\in\Omega}\left|(\mathcal{N}_E * K)(\bm{y})\right|}
\end{equation}
\begin{equation}
\widehat{\mathcal{N}}_S(\bm{x}) = \frac{(\mathcal{N}_S * K)(\bm{x})}{\displaystyle \max_{\bm{y}}\left|(\mathcal{N}_S * K)(\bm{y})\right|}
\end{equation}

Accordingly,
\begin{equation}
\widehat{\mathcal{N}}_E(\bm{x}) \in [-1,1]
\qquad
\widehat{\mathcal{N}}_S(\bm{x}) \in [-1,1]
\end{equation}

The stochastic fields are then constructed using a symmetric log-scale multiplicative model. For the \textit{E. coli} field, prescribe the half-span of the base-10 logarithmic multiplier $\alpha_E$ is chosen as 2.0, and define the multiplicative factor
\begin{equation}
M_E(\bm{x}) = 10^{\alpha_E\widehat{\mathcal{N}}_E(\bm{x})} \in \left[10^{-2}, 10^2\right]
\end{equation}
The preliminary stochastic \textit{E. coli} field is then
\begin{equation}
E^{(0)}(\bm{x}) = E_{{base}}(\bm{x}) M_E(\bm{x})
\end{equation}

Similarly, for the salt field, we choose a smaller $\alpha_S = 0.1$ considering the factor that the salt solute is non-active and thus distribute more in align with the background fluid field. Then,
\begin{equation}
M_S(\bm{x}) = 10^{\alpha_S\widehat{\mathcal{N}}_S(\bm{x})} \in \left[10^{-0.1}, 10^{0.1}\right]
\end{equation}
and the preliminary stochastic salt field is
\begin{equation}
S^{(0)}(\bm{x}) = S_{{base}}(\bm{x}) M_S(\bm{x})
\end{equation}

We construct the random multiplier $\alpha_E$ and $\alpha_S $ in logarithmic rather than linear scale by assuming equal
relative (fold-change) rather than equal absolute departures from the base field. 

For the salt-only control case, the \textit{E. coli} profile is set to zero before the stochastic mapping:
\begin{equation}
\tilde{e}(r) \equiv 0.
\end{equation}

For bacteria-coupled cases, the preliminary salt field is further modulated by a bacteria co-localization bias $B(\bm{x})$ considering the fact that \textit{E.~coli} tends to be the sites that accumulate salt ions:
\begin{equation}
B(\mathbf{x}) = \eta + (1-\eta)\,E_{{ctrl}}(\mathbf{x})^{\gamma}
\end{equation}
where $\eta \in [0, 1]$ sets the salt-following-bacteria coupling strength
($\eta=1$ recovers an \textit{E. coli}-independent salt field) and $\gamma>0$
controls the nonlinear emphasis of \textit{E. coli}-rich regions. We used $\eta = 0.1$ and $\gamma = 2.0$ in this modeling. $E_{ctrl}$ is a normalized log-compressed measure of how bacteria-rich a location is relative to the rest of the domain, scaled to $[0,1]$:

\begin{equation}
E_{{ctrl}}(\bm{x}) =
\frac{\log\!\left(1+\dfrac{E_{{stoch}}(\bm{x})}{\bar{E}_{{ref}}}\right)}
{\displaystyle\max_{\bm{y}\in\Omega}
\log\!\left(1+\dfrac{E_{{stoch}}(\bm{y})}{\bar{E}_{{ref}}}\right)}
\;\in [0,1]
\end{equation}
where $\bar{E}_{{ref}}$ denotes the median of $E_{{stoch}}(\bm{x})$
over $\Omega$.

The salt field is then updated as
\begin{equation}
S^{(0)}(\bm{x}) \leftarrow S^{(0)}(\bm{x})\,B(\bm{x})
\end{equation}
For the salt-only control case, the co-localization bias is disabled by setting $B(\bm{x}) \equiv 1$.

Both stochastic fields are subsequently rescaled to preserve total substrate-integrated mass. Denoting the final mass-conserved fields by $E_{{stoch}}(\bm{x})$ and $S_{{stoch}}(\bm{x})$, one has
\begin{equation}
E_{{stoch}}(\bm{x}) = E^{(0)}(\bm{x}) \frac{\displaystyle\int_{\Omega} E_{{base}}(\bm{y}) \, dA}{\displaystyle\int_{\Omega} E^{(0)}(\bm{y}) \, dA}
\end{equation}
\begin{equation}
S_{{stoch}}(\bm{x}) = S^{(0)}(bm{x}) \frac{\displaystyle\int_{\Omega} S_{{base}}(\bm{y}) \, dA}{\displaystyle\int_{\Omega} S^{(0)}(\bm{y}) \, dA}
\end{equation}

\subsubsection{Nucleation-Limited Precipitation Model}

From the final stochastic fields \(S_{{stoch}}(\bm{x})\) and \(E_{{stoch}}(\bm{x})\), we choose the baseline absolute nucleation threshold as $S_0 = 0.1~mol/m^2$. For the salt-only control, \(S_{{crit}}(\bm{x})=S_0\).  
For bacteria-coupled cases, threshold lowering is driven by a smooth logarithmic response to absolute local \textit{E. coli}:
\begin{equation}
\Phi_E(\bm{x})=1-\bigg(1+\frac{E_{stoch}(\bm x)}{E_{ref, abs}}\bigg)^{-\beta}
\end{equation}
\begin{equation}
S_{{crit}}(\bm{x})=
S_0\,[1-\Phi_E(\bm{x})]
\end{equation}

Here \(\Phi_E(\bm{x})\in[0,1)\) is a dimensionless bacterial threshold-lowering factor: it equals \(0\) where no bacteria are present, and approaches \(1\) as local \textit{E. coli} density grows without bound, so \(S_{{crit}}(\bm{x})\) is pulled continuously from \(S_0\) toward \(0\) as bacteria accumulate at a location. The reference density \(E_{{ref,abs}}\) sets the characteristic \textit{E. coli} density at which threshold-lowering becomes significant, while \(\beta\) controls how sharply \(\Phi_E\) rises once local density approaches \(E_{{ref,abs}}\). We use \(\beta=1.6\) and \(E_{{ref,abs}}=5\times10^{10}\ {cells/m^2}\), a typical areal density close to the interior region density seen in our tested cases (see Fig.~\ref{fig:pattern_modeling}).

Primary nucleation is defined by
\begin{equation}
P_0(\bm{x})=
\mathbf{1}\!\left[S_{{stoch}}(\bm{x})>S_{{crit}}(\bm{x})\right]
\end{equation}
where $\mathbf{1}[\cdot]$ means that the function returns "1" (crystallized) if the condition $[\cdot]$ is satisfied.

Once a crystal forms, it is easier for salt to nucleate right next to it, since the existing crystal surface locally lowers the energy barrier for further growth. We capture this "neighbor-assisted" effect in three steps. First, we smooth the current nucleation pattern \(P(\bm{x})\) with a local averaging kernel \(K_n\) (for simplicity, chosen the same as $\bm K$ in Eq.~\ref{eq:kernel}) and rescale it to the range \([0,1]\), giving a local nucleation density \(N(\bm{x})\): a value near 1 means a location is surrounded by many already-nucleated neighbors, while 0 means it has none.
\begin{equation}
N(\bm{x})=\frac{(K_n*P)(\bm{x})}{\max_{\Omega}(K_n*P)}
\end{equation}
Second, this local density lowers the nucleation threshold near existing crystals $S_{{crit,nb}}$, with strength set by \(\mu\in[0,1]\): a location fully surrounded by nucleated neighbors (\(N=1\)) has its threshold reduced by a factor of \(\mu\), while an isolated location (\(N=0\)) keeps its original threshold.
\begin{equation}
S_{{crit,nb}}(\bm{x})=
(S_{{crit}}(\bm{x})\,[1-\mu  N(\bm{x})]
\end{equation}
Third, any location that now clears this relaxed, neighbor-lowered threshold is added to the nucleation mask:
\begin{equation}
P(\bm{x})\leftarrow
P_0(\bm{x})\lor
\mathbf{1}\!\left[S_{{stoch}}(\bm{x})>S_{{crit,nb}}(\bm{x})\right]
\end{equation}
We use \(\mu=0.75\), giving strong neighbor-assisted expansion.

Finally, precipitated salt is assigned directly on nucleated pixels,
\begin{equation}
S_{{precip}}^{(0)}(\bm{x})=P(\bm{x})\,S_{{stoch}}(\bm{x})
\end{equation}
and total salt is conserved by uniform rescaling:
\begin{equation}
S_{{precip}}(\bm{x})=
S_{{precip}}^{(0)}(\bm{x})
\frac{\int_{\Omega}S_{{stoch}}\,dA}
{\int_{\Omega}S_{{precip}}^{(0)}\,dA}
\end{equation}
This formulation predicts where nucleation and precipitation are favored from transport-limited fields with explicit bacteria-coupled threshold modulation and strict mass conservation, while not resolving microscale crystal morphology.

\clearpage
\bibliographystyle{elsarticle-num}
\bibliography{references}
\end{document}